\documentclass[11pt,a4paper,onecolumn]{IEEEtran}

\usepackage[dvipsnames]{xcolor}
\usepackage[utf8]{inputenc}  
\usepackage{url}
\usepackage{amsmath}
\usepackage[makeroom]{cancel}
\usepackage{amsfonts}
\usepackage{amssymb}
\usepackage{booktabs}
\usepackage[pdftex]{graphicx}
\usepackage[table]{xcolor}
\graphicspath{{../pdf/}{../jpeg/}}
\usepackage{pgfplots}
\pgfplotsset{ 
  compat=newest, 
}
\usepgfplotslibrary{fillbetween}
\usetikzlibrary {positioning}
\DeclareGraphicsExtensions{.pdf,.jpeg,.png}
\usepackage{tikz}
\usepackage[left=2.25cm,right=2.25cm,top=2cm,bottom=2cm]{geometry}

\author{Mengbin Ye$^{1,2,\star}$, Wooseok Jung$^{2}$, Tony J. Mathew$^{2}$, Lorenzo Zino$^3$, and Yoshihisa Kashima$^4$\\ \vspace{11pt}
\small{$^1$Adelaide Data Science Centre, University of Adelaide, Adelaide, Australia\\
$^2$Centre for Optimisation and Decision Science, Curtin University, Perth, Australia\\
$^3$Department of Electronics and Telecommunications, Politecnico di Torino, Torino, Italy \\
$^4$School of Psychological Sciences, University of Melbourne, Melbourne, Australia\\
$^\star$Corresponding author: e-mail: ben.ye@adelaide.edu.au  }}

\title{\LARGE \textbf{An agent-based model of the formation and evolution of common ground}}
\usepackage{tabularx}
\usepackage{float}
\usepackage{amsthm}  
\usepackage{enumitem} 
\usepackage{bbm} 
\usepackage{cite}
\usepackage{graphicx}
\usepackage{subfig}

\renewcommand{\eqref}[1]{Eq.~(\ref{#1})}  

\newcommand{\vertiii}[1]{{\left\vert\kern-0.25ex\left\vert\kern-0.25ex\left\vert #1 
    \right\vert\kern-0.25ex\right\vert\kern-0.25ex\right\vert}} 

\date{}

\usepackage{hyperref} 
\hypersetup{
	citebordercolor = 1 1 1,
	linkbordercolor = 1 1 1,
	menubordercolor = 1 1 1,
}

\usepackage{cleveref}

\begin{document}



\maketitle

\begin{abstract}
The existence of a communal common ground is vital for collective action and coordination in a population, but the micro-level cognitive and social processes that lead to the formation and evolution of common ground at the macro-level are undertheorised and have not been rigorously explored. In this work, we adopt a formal approach and develop an agent-based model that describes repeated grounding attempts between agents interacting on a network, with an explicit distinction between a sender agent and a receiver agent during an interaction involving sharing information. Several key novel features enable us to capture a range of different interaction contexts: we allow for the interaction to result in either acceptance or rejection, the receiver’s response may be lost to the sender, and the sender can interpret this lack of response as either acceptance or rejection (or even something in between). A campaign of Monte Carlo simulations reveals how different interaction contexts, as well as the available information for sharing, result in different emergent phenomena, such as a global communal common ground, fragmentation into multiple clusters of differing common ground, and even the total loss of any shared common ground.  This work highlights the potential for using mathematical models to study micro-macro links in cultural dynamics, including identifying ways to facilitate interactions to foster the emergence of a global communal common ground.
\end{abstract}


\section{Introduction}\label{sec:intro}

Coordination of individuals' actions for collective benefit has been and still is critical for humanity's success as a group-living species~\cite{de2019common,colman2014explaining,kashima2023sustainability,Coleman2021}. From hunting large game to recovering from a natural disaster, achieving a goal that an individual cannot achieve on their own requires a group to coordinate their actions to reap the benefits of the group goal. In coordinating everyone's actions, predictions, and expectations, the individuals' shared understanding of each other's actions is crucial. A condition necessary for coordination is called \textit{common knowledge} (e.g., \cite{schelling1957bargaining, thomas2014psychology,de2019common,chwe2013rational,rubinstein1989electronic,clark1996using,shteynberg2020shared}). It should not be confused with shared knowledge, which each person privately knows, but does not know that others do. In contrast, common knowledge requires that not only everyone knows, but everyone knows that everyone knows. As Steven Pinker \cite{pinker2025everyone} recently reminded us, that the emperor had no clothes was merely shared knowledge to the townsfolk in the famous Hans Christian Andersen's tale. It became common knowledge only when an innocent child blurted out ``The emperor has nothing on!" 

Common knowledge is a cognitively puzzling phenomenon at one level. It involves an infinite chain of recursion -- everyone knows that everyone knows that everyone knows...\textit{ad infinitum}... how to act in a group. However, a finite human mind cannot handle an infinite recursion (e.g., \cite{lewis1969convention,shteynberg2020shared,schelling1957bargaining,clark1996using,pinker2025everyone}). How is the common knowledge problem resolved in finite human minds in a group, especially in a large collective involving a large number of people? Solving this puzzle is all the more important in the contemporary world, where large-scale coordination is needed to address challenges facing humanity, including climate change, pandemics, and geopolitical instability, amid increasing signs of political polarization in many democracies worldwide~\cite{westfall2015perceiving,judge2023environmental}.

In the present article, we argue that one way of resolving the puzzle of common knowledge in a large collective is to construe it through the lens of \textit{common ground} and \textit{grounding}~\cite{clark1991grounding,clark1996using}, and provide a proof of concept by developing an agent-based model of grounding that can explain the emergence of a large-scale communal common ground through repeated interactions between agents (representing people). In what follows, we first clarify our terminology, such as common knowledge and common ground,  describe the implications of this analysis for contemporary cultural dynamics, highlighting the contrast between face-to-face and online social contexts. We then introduce our grounding model, and show its implications, suggesting that online contexts are particularly likely to fragment common ground and disrupt the coordination within a large-scale collective.

\subsection*{Shared knowledge, common knowledge, and common ground}

The nomenclature of shared knowledge, common knowledge, and common ground is used in various ways, and different authors use them somewhat differently. In this context, ``knowledge" is used in a broad sense, including knowledge, beliefs, and other forms of cognitive representation or ideational content. We use the term \textit{shared knowledge} to refer to the knowledge that each individual has just as others do (also \cite{thomas2014psychology}). To use a concrete example, imagine two individuals, Alice (\textit{A}) and Bob (\textit{B}), sitting together and a dark cloud is approaching them. We call this situation the state of affairs $s$. If each of (\textit{A}) and (\textit{B}) has certain knowledge $p$ (e.g., ``A rain cloud is coming"), $p$ is shared knowledge of (\textit{A}) and (\textit{B}). In contrast, \textit{common knowledge} has an additional property. Each person needs to know that everyone else knows that everyone else knows that... \textit{ad infinitum}...  $p$ (also \cite{vanderschraaf2023stanford,shteynberg2020shared,thomas2014psychology}). 
In other words, common knowledge involves an infinite chain of recursion, a condition logically necessary for coordinating actions among individuals and coordinating everyone's others' action predictions and expectations (e.g., \cite{schelling1957bargaining, thomas2014psychology,de2019common,chwe2013rational,rubinstein1989electronic,clark1996using}).

\textit{Common ground}, as we use the term with Herbert Clark \cite{clark1996using}, refers to a ground or a basis on which a truncated chain of recursion in common knowledge may be inferred by everyone involved. Somewhat more precisely, knowledge \textit{p} is common ground for everyone if and only if:
\begin{enumerate}
    \item everyone has information that \textit{s} holds;
    \item \textit{s} indicates to everyone that everyone has information that \textit{s} holds; and
    \item \textit{s} indicates to everyone that \textit{p}.
\end{enumerate}

Herbert Clark (\cite[p. 94]{clark1996using}) called this Common Ground (shared basis), or CG-shared, and acknowledged its providence in David Lewis's discussion of common knowledge~\cite[pp. 52--60]{lewis1969convention}. As Lewis noted, this inference about $p$ depends on ``suitable ancillary premises" (p. 53), such as a certain inductive reasoning capacity, inductive standards, and background information. The result may be at most several steps in the chain of recursion, e.g., (\textit{A}) knows that (\textit{B}) knows that $p$, achieving what a cognitive hierarchy model in behavioural game theory assumes in strategic social interaction (\cite{camerer2004cognitive,camerer2015psychological}). In other words, common ground \textit{p} is a result of bottom-up perceptual-cognitive processes based on a state of affairs \textit{s} in the world. This understanding of common ground broadly overlaps with what Shteynberg and his colleagues \cite{shteynberg2020shared} called the collective attention solution to the problem of infinite recursion in common knowledge.


It is important to note that the state of affairs \textit{s} that serves as a basis of common ground is not always part of the environment independent of the agents, such as the dark cloud in the sky. It could be the agents' behaviours and their effects in the world. Consider the following exchange between Alice and Bob:

\begin{itemize}
    \renewcommand{\labelitemi}{}
    \item A: You know that building wall with a picture on it...
    \item B: Ah, the street art near the station...
    \item A: Yes, Sue --- that building's owner said she was going to paint over it... real pity...
\end{itemize}

\noindent This would establish common ground: the picture on the building wall near the station is a piece of street art. In this social interaction, initial common ground is established (i.e., that building wall near the train station), and new common ground is subsequently added (i.e., there is a piece of street art on the building wall). In this example, the momentarily and linguistically constructed intersubjective object, ``that building wall," to which Alice and Bob's collective attention is directed, is the state of affairs \textit{s} serving as a basis for the common ground \textit{p}: what is on that building wall is street art. 

Clark and his colleagues called the process of creating and adding new common ground \textit{grounding}~\cite{clark1996using,clark1991grounding}. It consists of a communication \textit{sender's} behaviour that \textit{presents} a certain idea and a communication \textit{receiver's} behaviour that \textit{accepts} it. Presentation can be an utterance, or even a gesture (e.g., pointing to the dark cloud); acceptance can also take a variety of forms, insofar as it can provide sufficient evidence for the sender that the receiver has understood what the sender meant. Thus, to ground something is to exchange a \textit{presentation} and an \textit{acceptance}, allowing the establishment of  common knowledge as part of common ground well enough for current purposes of social interaction (e.g., Clark, 1996, p. 221). 

In the above example, (\textit{A})'s first utterance is a presentation, and (\textit{B})'s utterance accepts it, creating ``that building wall" as an object of their joint attention. At the same time, (\textit{B})'s utterance acts as a presentation of the new idea that the picture on that building wall is street art. The word ``graffiti" or ``vandalism" was not used instead of ``street art", allowing (\textit{B}) to present a certain understanding of the picture on the wall. By saying ``Yes" and continuing on to say that it is a pity to paint over it, (\textit{A}) tacitly accepts (\textit{B})'s presentation that the picture is street art. Note that (\textit{A}) could have \textit{rejected} (\textit{B})'s presentation by saying something like, ``Oh, you mean that graffiti?" or even calling it ``vandalism". Instead, (\textit{A}) tacitly ruled out those alternative potential interpretations, establishing the picture on the building wall as a work of art, rather than a mere graffiti, let alone an act of vandalism. 

The presentation-acceptance pairing in grounding may be construed as a tacit joint commitment \cite{gilbert2017joint} and act as an implicit ``pact" \cite{brennan1996conceptual}. Therefore, these agents would, and also are mutually entitled and obligated to, expect the common ground they have established in a previous interaction to continue to exist when they meet next time (within a reasonable limit, e.g., unless a very long time has elapsed), and they can coordinate their actions based on this common ground \cite{kashima2014meaning}. In this sense, grounding can create what Higgins and his colleagues called \textit{shared reality} \cite{echterhoff2009shared,higgins2021sr} (also see \cite{kashima2014meaning,kashima2018social} for further discussion).

\subsection*{Communal common ground} 

So far, we have discussed how two or more relatively small numbers (e.g., $\textit{n} \leq $5; see \cite{fay2000group}) of agents create and expand common ground. Nevertheless, a far greater number (e.g., $\textit{n} \gg $5) of agents can have common ground at a macro-level. For example, it is safe to say that a vast majority, if not all, of the citizens of the United States would have ``Donald Trump is the current US president as of 2026" in their common ground. Here, more than three hundred million residents of the United States constitute a \textit{cultural community} in which each agent would be willing to say not only do they know who the president is, but also that everyone would know who the president is, etc. Not only the citizens of a nation-state as an ``imagined community" \cite{anderson1983imagined}, but also other groups of agents can be regarded as a cultural community, insofar as they can coordinate their actions while engaging in a joint activity (e.g., \cite{scheff1967consensus}). They include a linguistic community (e.g., English speakers), a community of experts (e.g., ophthalmologists), and the like (see \cite{clark1996using}). The common ground of a cultural community is \textit{communal common ground} \cite{clark1996using}.

Communal common ground can arise through top-down and bottom-up processes. When a central authority issues a public announcement within a cultural community (e.g., an electoral commission declaring an election winner), it can establish new common ground \textit{top-down}. In contrast, agents can build up a communal common ground \textit{bottom-up} through numerous social interactions while engaging in repeated grounding processes \cite{kashima2007grounding,kashima2014meaning}. Imagine that after Alice and Bob's conversation about the street art, Bob may repeat a similar conversation with Catherine, who then talks about it with David, and so on. This way, even a cultural community without a central authority can have a bottom-up communal common ground; a cultural community with a central authority can have a communal common ground on a topic on which the central authority makes no declaration. 

Indeed, residents of a nation-state have some ideas about how most people in the country think, feel, and act (i.e., a topic that a central authority is unlikely to dictate), which can explain cultural differences in values, beliefs, and cognition between nation-states (e.g., \cite{fischer2006congruence,wan2007perceived,zou2009culture}). The communal common ground of a large-scale cultural community, such as a nation-state, has been termed \textit{intersubjective culture} \cite{chiu2010intersubjective}, which helps members of the community solve recurrent local coordination problems efficiently (e.g., \cite{chwe2013rational,kashima1999culture}). Nevertheless, not every population develops a \textit{global} communal common ground for the entire population under all circumstances. Depending on the micro-level grounding process, \textit{fragmented} communal common ground may emerge, in which multiple cultural communities exist with distinct communal common grounds. In an extreme case, there may be no communal common ground at all, arguably resulting in a state of \textit{anomie} \cite{scheff1967consensus}. 

\subsection*{Sender--Receiver Grounding model}

In the present article, we develop an agent-based model to provide a proof of concept that communal common ground, even at a large scale, can emerge through repeated grounding processes in piecemeal dyadic or small-group interactions among agents within a population, generating an intersubjective culture. We refer to the model as the Sender--Receiver Grounding (SRG) model, as its distinctive feature is the micro-level psychological processes of \textit{sender presentation} and \textit{receiver acceptance} in the structuration of communal common ground. In particular, we suggest that, depending on how a receiver responds to a sender's presentation of information, and how the sender interprets it, different macro-level communal common ground emerges, ranging from global to fragmented to anomic. Recall that grounding is successfully completed when a receiver receives information presented by a sender and the receiver accepts it by providing evidence sufficient for the sender to be satisfied with the receiver's understanding. In other words, a sender presenting information is insufficient for the information to be grounded --presented information is not grounded unless the sender receives sufficient evidence of the receiver's acceptance, either (i) in the receiver's response or (ii) by interpreting the absence of a response as acceptance. We will show whether communal common ground emerges, and what type of macro-level community structure emerges, critically depends on a receiver's response. This feature distinguishes the present model from other models of cultural dynamics (see for a review, \cite{kashima2017modeling}), such as Axelrod's model of cultural dissemination \cite{axelrod1997dissemination} and dynamic social impact theory \cite{latane1996dynamic}. 

Furthermore, the model sheds light on how the context of social interaction, particularly online social interaction, may unite or fragment communal common ground. Across different interaction contexts, the likelihood that a sender receives a receiver's response and how the sender interprets the absence of a receiver's response could vary substantially. In a dyadic or small-group face-to-face interaction, it is highly likely that a receiver provides some response to signal their acceptance of the presented information. In fact, the absence of a receiver response (i.e., silence) disrupts grounding \cite{benus2011pragmatic,brennan1995feeling} and is likely to be interpreted as a rejection of the presented information (\cite{kouden2021norm}. 
By contrast, in online social interaction (for a review, see \cite{van2021social}), the likelihood that a sender's presentation receives an acceptance response is much lower. Furthermore, the absence of a response may be interpreted by the information sender differently, as acceptance in some cases or rejection in others. How a sender interprets the absence of a response depends on a variety of factors, including the sender's proclivities (e.g., trusting) and the receiver's characteristics (e.g., receptive or hostile audience; \cite{echterhoff2005audience}).

We use numerical simulations to examine the link between micro-level grounding and macro-level communal common ground. First, we illustrate that both a global communal common ground and a fragmented common ground are emergent phenomena in our model. In the latter, there are multiple clusters of agents; a common ground is shared by agents within the same cluster, while different clusters hold different common ground. 
In our main study, we employ 
Monte Carlo simulations to establish clear insights into how the interaction context, together with the diversity of alternative information that can be selected for presentation in grounding, shapes the particular emergent macro-level phenomena, looking both at the transient temporal evolution and the long-term stability of the communal common ground. A key insight is that as the number of selectable information increases, fragmentation of the common ground becomes likely, especially if the sender is likely to perceive a receiver's response. However, certain information contexts will lead to a global communal common ground, irrespective of how many pieces of information are available for presentation. In particular, environments that favour a global communal common ground are those in which i) it is likely for the sender not to perceive the receiver's response, and ii) the lack of response is interpreted as rejection.  When the lack of a response is instead interpreted as acceptance, combined with a large number of information to select from, we instead see a collapse of the common ground; most agents have entirely unrelated perceptions of the common ground, resulting in an emergent state of anomie. The ability to, in a unified formal modelling framework, clearly relate the micro-level cognitive and social mechanisms of grounding to the macro-level cultural dynamics of common ground formation and evolution, underpins the main contribution of our work. 

The rest of the paper is structured as follows. In Section~\ref{sec:model}, we present the model and explain how the mathematical abstraction represents the grounding process. Section~\ref{sec:emergent_phenomena} presents several exemplar simulations to illustrate the different emergent phenomena. The main simulations and results are reported in Section~\ref{sec:results}, with discussions and conclusions in Section~\ref{sec:discussion}.

\section{Model}\label{sec:model}

\begin{figure}[!hb]
    \centering
    \includegraphics[width=\linewidth]{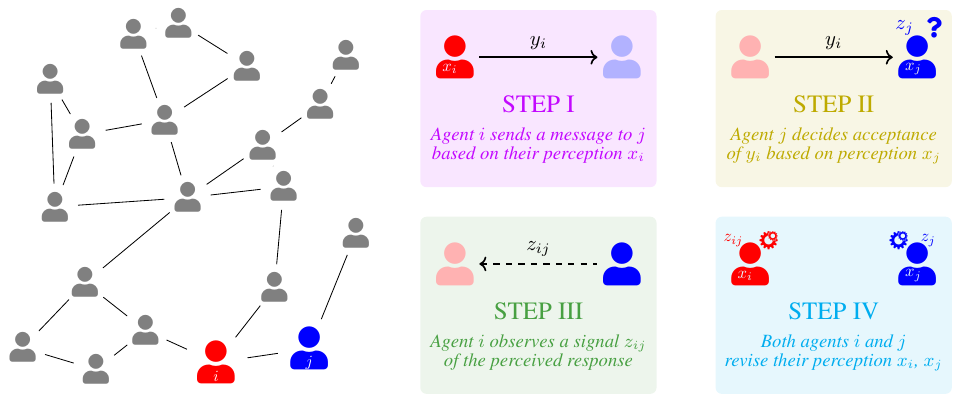}
    \caption{A schematic of the model. On the left, a population of agents interacting over a backbone network. At each time step, one agent becomes the sender (agent $i$, in red) and the sender selects one network neighbour to be the receiver (agent $j$, in blue) ---refer to Section~\ref{ssec:agents} for details. The right panel shows the four steps that occur in the grounding process ---refer to Section~\ref{ssec:info_transmission_model} for details. }
    \label{fig:schematic}
\end{figure}

In this section, we present the Sender--Receiver Grounding (SRG) mathematical model that aims to capture the grounding process, and through it, the formation of common ground among a group of individuals. We first describe the agents, their states, and the interaction mechanism. Then, we detail the information transmission process that plays out over any given interaction. A schematic illustrating the network model and the communication process is reported in Fig.~\ref{fig:schematic}.

\subsection{Agents, states, and interaction mechanisms}\label{ssec:agents}

We consider a population of $n \geq 2$ agents (individuals), which we index using positive integer numbers, i.e, $\mathcal{V} = \{1,2,\dots,n\}$. Agents interact with one another, sending information and receiving a response, over discrete time steps $t=0,1,2,\dots$. These interactions occur on an undirected unweighted graph $\mathcal{G}=(\mathcal{V},\mathcal{E})$, where $\mathcal{V}$ is the set of nodes (and thus each node corresponds to an agent) and $\mathcal{E}\subseteq \mathcal{V}\times\mathcal{V}$ is the unordered set of edges, whereby the presence of edge $(i,j)\in\mathcal E$ means that agents $i$ and $j$ can interact. We use $\mathcal{N}_{i} = \{j\in\mathcal V:(i,j) \in \mathcal{E}\}$ to denote the neighbour set of agent $i$, i.e., all the agents with whom $i$ can interact. See the left panel of Fig.~\ref{fig:schematic} for an illustrative example of this network.


In the SRG model, information exchange occurs between agents on a single cultural feature (e.g., slang used in online conversations), of which there are $m\geq 2$ pieces of mutually exclusive information to select from (e.g., rizz, brain rot, aura farming, ...). For convenience, we define the set of information comprising the cultural feature as $\mathcal{I} = \{1, 2, \hdots, m\}$. Each agent $i\in\mathcal V$ has a state vector $x_{i}(t)=[x_{i,1}(t),x_{i,2}(t),\dots,x_{i,m}(t)]^{\top} \in [-1,+1]^{m}$, which represents agent $i$'s perception of the common ground.  
We interpret $x_{i,k} \in [-1,+1]$ as agent $i$'s level of certainty that information $k$ is known by other agents, i.e. belongs to the common ground, where $k \in \mathcal{I}$. That is, $x_{i,k}=-1$ and $x_{i,k}=+1$ correspond to agent $i$ being maximally certain that information $k$ is not known to other agents (not part of the common ground) and is known to other agents (part of the common ground), respectively. Meanwhile, $x_{i,k}=0$ means that agent $i$ is neutral about whether $k$ is part of the common ground. Through this framing, we interpret $x_i(t)$ to be agent~$i$'s representation of the common ground, and in particular what information is and is not part of said common ground. Each agent's state evolves through repeated encounters with other agents, information transmission, and the associated success or failure to ground in that encounter, as detailed in the following. 

Agents interact over discrete time steps, and each agent can act as a \textit{sender} or a \textit{receiver}. We employ a standard asynchronous mechanism that is often used in agent-based modelling. In this scheme, at each time step $t$ and independent of past events, a single agent is selected uniformly at random from the population, $\mathcal{V}$, and assigned to be the {sender}. Supposing that agent~$i$ is the sender at time $t$, then agent~$i$ will select a single agent $j$ uniformly at random from their set of neighbours, $\mathcal{N}_{i}$, to act as a receiver. In other words, at every time step, a single pair of agents interact, one as the sender and the other as the receiver. This gossip-like asynchronous dynamics allows for time-varying interactions that better reflect how conversations occur and information is exchanged, but is constrained by the backbone network $\mathcal G$. See the left panel of Fig.~\ref{fig:schematic}, in which the sender and receiver are highlighted in red and blue, respectively, while other agents are grey and inactive.

\subsection{Information transmission and grounding process}\label{ssec:info_transmission_model}

Once the sender and receiver roles have been determined, the grounding process plays out in four steps. Assuming that a given agent~$i$ is the sender and another given agent~$j$ is the receiver, we now give a summary of the process before presenting the mathematical implementation. In Step~I, the sender, agent $i$, selects a piece of information $k \in \mathcal{I}$ to communicate to the receiver, agent $j$. For instance, the sender inserts the phrase ``aura farming'' into a comment they send to the receiver. For Step~II, the receiver (agent $j$) decides whether to accept or reject the communication. Step~III completes the interaction process, as agent~$j$ (possibly) sends a signal back to agent $i$ about the outcome (in our example, the receiver sends a reply that signifies they did or did not understand the comment, or sends no reply). In Step~IV, based on the outcome of the interaction, the perceptions of the common ground of both agents, viz. $x_i(t)$ and $x_j(t)$ are updated. This concludes the grounding process at time $t$. These four steps, illustrated in the right panel of Fig.~\ref{fig:schematic} with the sender in red and receiver in blue, are detailed in the following. For a summary of all variables and parameters involved in the dynamics, we refer to Table~\ref{tab:parameters}.

\begin{table}    \caption{Model setting (red), variables (blue), and parameters (purple). }
\definecolor{cifRobinsEggBlue}{RGB}{0,153,153} 
\definecolor{cifElectricPurple}{RGB}{168,51,255}
\definecolor{cifFreeSpeechRed}{RGB}{174,0,2}
\definecolor{cifSherpaBlue}{RGB}{0,76,76}
\definecolor{cifCoral}{RGB}{255,135,63}
    \label{tab:parameters}
    \centering
\begin{tabular}{r|l}
\rowcolor{cifFreeSpeechRed!10}$\mathcal V=\{1,\dots,n\}$& Set of agents (individuals). Agent~$i$ refers to a generic agent in the population\\
\rowcolor{cifFreeSpeechRed!5}$\mathcal I=\{1,\dots,m\}$& Set of information. Information~$k$ refers to a generic piece of information\\
\rowcolor{cifFreeSpeechRed!10}$\mathcal N_i\subseteq \mathcal V$& Set of neighbours of agent $i$\\
\hline
\rowcolor{cifSherpaBlue!10}$x_{i,k}(t)\in[-1,+1]$& Agent~$i$'s level of certainty that information $k$
belongs to the common ground, at time $t$\\
\rowcolor{cifSherpaBlue!5}$y_i(t)\in\mathcal I$& The information selected by agent $i$ to transmit at time $t$\\
\rowcolor{cifSherpaBlue!10}$z_j(t)\in\{-1,+1\}$& Acceptance ($z_j=1$) or rejection ($z_j=-1$) of the information by the receiver agent $j$ at time $t$\\
\rowcolor{cifSherpaBlue!5}$z_{ij}(t)\in\{-1,0,+1\}$& Acknowledgment of acceptance/rejection sent by agent $j\in\mathcal V$ to $i\in\mathcal V$ at time $t$\\\hline
\rowcolor{cifElectricPurple!10}$\beta_i\in[0,\infty)$& Selection bias of agent $i\in\mathcal V$\\
\rowcolor{cifElectricPurple!5}$\varepsilon\in[0,1]$& Probability of signal loss: No explicit response from the receiver agent reaches the sender agent\\
\rowcolor{cifElectricPurple!10}$\gamma\in[-1,+1]$& Interpretation bias: How the sender agent interprets the lack of an explicit response\\
\rowcolor{cifElectricPurple!5}$\alpha_i\in[0,1]$& Susceptibility of agent $i\in\mathcal V$ to social influence\\
\rowcolor{cifElectricPurple!10}$\sigma\in[0,1]$& Decay factor    \end{tabular}
\end{table} 

{\bf Step I.} Let us define $y_{i}(t) \in \mathcal{I}$ as the information that is transmitted by agent $i$ at time $t$. Agent~$i$'s decision on which information to select depends on their perception of the common ground, and is determined by a logit response rule (also referred to the softmax function in machine learning or log-linear learning in game theory)~\cite{lewandowsky2010computational,blume1995best_response}. Specifically, the probability that agent~$i$ selects information~$k$ at time $t$  is equal to
\begin{equation}\label{eq:sender_decision}
    \mathbb{P}[y_{i}(t)=k]= \frac{e^{\beta_{i}x_{i,k}(t)}}{\sum_{l=1}^{m} e^{\beta_{i}x_{i,l}(t)}}.
\end{equation}
Here, $\beta_i \in [0,\infty)$ is a parameter that shapes agent $i$'s selection process, conveniently referred to as the selection bias, whereby $\beta_i = 0$ corresponds to choosing $y_{i}(t)=k$ uniformly at random, and $\beta_i \to \infty$ corresponds to strictly choosing the information that they are most certain belongs to the common ground.

{\bf Step~II.} We use $z_{j}(t)\in\{ -1,+1 \}$ to record whether agent $j$ accepts or rejects the transmitted information $y_{i}(t)=k$, where $z_{j}=+1$ and $z_{j}=-1$ correspond to acceptance and rejection, respectively. Recalling that $x_{j,k}$ is the certainty agent~$j$ has of information~$k$ being in the common ground, we posit that the acceptance probability of the transmitted information, $y_i(t)$, is governed by a logistic function:
\begin{equation}\label{eq:receiver_acceptance}
    \mathbb{P}[z_{j}(t)=+1]=\frac{1}{1+e^{-\beta_{j}x_{j,y_{i}(t)}(t)}}.
\end{equation}
The inflection point is at $x_{j,y_{i}(t)}=0$, so if agent $j$ is neutral about the presence of information $y_i(t)$ in the common ground, they have a $50\%$ probability of accepting. The sharpness of the transition is governed by the selection bias $\beta_j$. \eqref{eq:receiver_acceptance} posits that agent~$j$'s probability of accepting or rejecting the transmission increases as $x_{j,y_i(t)}$, their certainty in the information belonging or not belonging to the common ground, increases or decreases, respectively. Given a fixed $x_{j,y_i(t)}$, increasing $\beta_j$ corresponds to a higher probability of accepting or rejecting, according to the sign of $x_{j,y_i(t)}$, i.e., whether it is positive or negative. 

{\bf Step~III.} After agent $j$ has decided to reject or accept the transmission from agent $i$, they attempt to communicate the outcome to agent~$i$. This process is captured by a variable $z_{ij}(t)$ generated according to the following equation:
\begin{equation}\label{eq:receiver_response}
    z_{ij}(t)=\begin{cases}
z_{j}(t) & \text{with probability } 1-\varepsilon, \\
\gamma & \text{with probability } \varepsilon.
\end{cases}
\end{equation}
Here, $\varepsilon \in [0,1]$ captures the probability that no explicit response is received by agent~$i$, while the constant $\gamma \in [-1,+1]$ captures how agent $i$ interprets the lack of response from agent $j$. For convenience, we will sometimes refer to  $\varepsilon$ as the probability of signal loss, and $\gamma$ as the interpretation bias. As we will explain in the sequel, $\gamma$ can represent different interaction contexts in which the lack of a response is treated differently by the sender. Note that this missing response could be due to a range of reasons, such as agent~$j$ not actually responding, or the response being lost in the process.

{\bf Step~IV.} With the transmission procedure complete, both agents update their perception of the common ground according to the following two equations:
\begin{subequations}\label{eq:agent_update}
\begin{align}
&x_{i,y_{i}(t)}(t+1)=(1-\alpha_{i})x_{i,y_{i}(t)}(t)+\alpha_{i}z_{ij}(t), \label{eq:sender_update}\\
&x_{j,y_{i}(t)}(t+1)=
\begin{cases}
    (1-\alpha_{j})x_{j,y_{i}(t)}(t)+\alpha_{j} & \text{ if } z_j(t) = +1 ,\\
    x_{j,y_{i}(t)}(t) & \text{ if } z_j(t) = -1.
\end{cases}\label{eq:receiver_update}
\end{align}
\end{subequations}
The parameter $\alpha_{i} \in [0,1], i \in \mathcal{V}$ can be thought of as the susceptibility of the agent $i$ to social influence, viz. how easily their level of certainty changes due to the interaction. From another perspective, $x_{i,y_i(t)} (t+1)$ is a weighted average (or convex combination) of the current certainty $x_{i,y_i(t)} (t)$ and $z_{ij}(t)$, being agent~$i$'s interpretation of the response from agent~$j$. Thus, the sender's certainty $x_{i,y_i(t)}(t)$ increases or decreases if they receive an acceptance or rejection signal from agent~$j$, and moves towards the value of $\gamma$ if no signal is received. The receiver $j$ on the other hand, increases their certainty $x_{j,y_i(t)}(t)$ if grounding attempt is successful ($z_j(t) = +1$) and makes no change if they rejected the grounding attempt. The certainties $x_{i,k}$ therefore remain between $-1$ and $+1$ for all time.



Moreover, decay occurs for the pieces of information that the receiver did not encounter:
\begin{equation}\label{eq:receiver_decay}
    x_{j,\ell}(t+1)=(1-\alpha_{j}\sigma)x_{j,\ell}(t)-\sigma\alpha_{j},
\end{equation}
for all $\ell \in \mathcal{I}$ with $\ell \neq y_{i}(t)$, where $\sigma \in [0,1]$ is the decay rate.
This concludes the model dynamics for time step $t$. 

\subsection{Key features of the SRG model}\label{ssec:motivation_model}


We now discuss some conceptual and theoretical underpinnings for the proposed mathematical model, as well as how it contrasts with existing models of cultural dynamics.

As discussed in the Introduction, existing literature has theorised that grounding is a key micro-level cognitive and social process for the establishing of a common ground between two individuals. The SRG model aims to represent this grounding process occurring at each time step as an interaction between two agents, with a clear sender (the one who initiates the process) and the receiver (the one who determines whether the process is successful or not). The formation and evolution of common ground at the population level necessarily arises from these repeated grounding attempts.

We also clearly separate the decision-making process as captured by \eqref{eq:sender_decision} and \eqref{eq:receiver_acceptance} (what information to send, and whether the receiver accepts or rejects the grounding attempt) from the updating of the agent states as a result of the interaction, viz. Eqs.~(\ref{eq:receiver_response})--(\ref{eq:receiver_decay}). Our approach is inspired by established models of noisy decision-making and selection processes~\cite{blume1995best_response,lewandowsky2010computational}, and the state updating dynamics follow classical models of weight averaging via social influence~\cite{proskurnikov2017tutorial,friedkin1990_FJsocialmodel}. Thus, the sender averages between their existing state and their understanding of the interaction outcome; the receiver does similarly when they accept the information, but makes no change when there is rejection. The decay in \eqref{eq:receiver_decay} represents a decrease in the receiver's certainty that other pieces of information (which they did not receive in the grounding attempt) belong to the common ground.

Another key feature of the SRG model is that the sender~$i$ makes a determination on what to send, $y_i(t)$, based only on their own perception of the common ground. That is, \eqref{eq:sender_decision} is a function of just $x_i(t)$. Similarly, receiver~$j$ decides to accept or reject based only on their certainty of the transmitted information, $x_{j,y_i(t)}(t)$, as in \eqref{eq:receiver_acceptance}. This contrasts established models including those based on coordination~\cite{ramazi2016networks} and synchronisation mechanisms~\cite{arenas2008synchronization}, in which agent~$i$ directly attempts to move its state $x_i(t)$ to match the state $x_j(t)$ of neighbour agent~$j$. It also contrasts other models of cultural dynamics and language evolution, including the seminal work by Axelrod~\cite{Axelrod1997} and the Naming Game model~\cite{baronchelli2008depth}, which take an almost evolutionary biological approach: $x_i(t)$ is more akin to a genetic trait, and when two agents~$i$ and $j$ interact, there is a probability of one agent adopting the trait(s) of the other agent as a function of $x_i(t)$ and $x_j(t)$. We argue that the clear separation of i) what information is sent, and ii) what information agents perceive to belong to the common ground, is what allows us to address the epistemic component of the common ground described in the Introduction. That is, we are specifically modelling the dynamics of whether people perceive that others share sufficiently similar knowledge. This shared knowledge becomes common knowledge when $x_i(t)$ and $x_j(t)$ are sufficiently ``close'', and thus a common ground is formed.

Finally, we expand on discussions presented in the Introduction, on how \eqref{eq:receiver_response} and the associated parameters $\varepsilon$ and $\gamma$, namely the response mechanism, can be used to conceptualise different communications or information environments that yield different interaction contexts. The loss of the response signal, governed by $\varepsilon$, might occur for different reasons. In a face-to-face conversation, agent~$i$ might say something to agent~$j$, who looks back blankly without saying anything. In an online chatroom, agent~$i$ might post a comment and even though agent~$j$ replies, agent~$i$ is unaware of the reply because they never checked back into the discussion thread. The probability the former occurs is substantially lower than the latter, and thus setting $\varepsilon = 0.1$ and $\varepsilon = 0.5$ can distinguish these scenarios. Meanwhile, the interpretation bias $\gamma$ represents how agent~$i$ interprets the lack of a response. In the face-to-face example, the lack of a response might be socially awkward (especially if $i$ and $j$ do not know each other well), causing agent~$i$ to doubt themselves and treat the lack of response as being similar to a rejection. This can be captured by $\gamma < 0$, with the magnitude indicating how similar it is to a true rejection. In online chatrooms, the lack of a response can occur with greater frequency and might feel less confronting. Thus even if the sender fails to receive a response, they may interpret this as a neutral outcome ($\gamma = 0$) or acceptance ($\gamma > 0$).

While the above are examples of different communication contexts and information environments, we note our model is an abstract representation of the interaction process, and thus the pair $\varepsilon$ and $\gamma$ can be used to represent a range of different real-world scenarios with differing probability and interpretation by the sender when failing to receive a response. Importantly, this mechanism means that we can distinguish between consistent and inconsistent understanding of the outcome of the grounding attempt. The case where $z_{ij}(t) = z_j(t)$ and $z_{ij}(t) = \gamma$ corresponds to there being consistency and inconsistency in said understanding. In Section~\ref{sec:discussion}, we discuss future directions that explore different implementation choices without substantially changing the framework of the proposed model. Note that we distinguish between the modelling framework (e.g., distinct decision-making processes for the sender and the receiver, and explicit acceptance or rejection mechanisms) and specific choices of implementation (e.g., pairwise asynchronous interactions, the receiver making no change when they reject the grounding attempt). 


\section{Emergent collective phenomena}\label{sec:emergent_phenomena}

Due to the complexity of the SRG model, this paper will use numerical simulations to characterise its dynamics. In this section, we provide exemplar simulations to showcase the diverse range of emergent phenomena that can arise, and discuss how this maps to different common ground formation processes. From these insights, we identify two core research questions that we address in the next section using a campaign of Monte Carlo simulations. 

\subsection{General simulation setup}\label{ssec:sim_setup}
Here, we describe the general simulation setup that is employed in this paper, clarifying several parameters and choices that are held constant across all simulations reported in this paper. Specific adjustments to the setup are described at the relevant points in the sequel. The complete set of Python code is available at the following Github repository: \url{https://github.com/anon-233131/SRG-2026}.

We consider a population of $n=500$ agents, which is sufficiently large so that a diverse range of population-level phenomena can arise, but not so large as to impose problems of computational complexity. 
The agents interact over an undirected, unweighted graph $\mathcal{G}$ which is a random small-world network generated using the Watts-Strogatz algorithm~\cite{newman2010networks_book}. 
This enables us to account for stochasticity, ensuring that our observations are robust, with each network having topological properties that real-world social networks are known to exhibit, viz. highly clustered communities with sparse weak ties between them, and a small-world effect. In Appendix~\ref{app:ws}, we provide a short description of the Watts-Strogatz algorithm, a brief discussion of its main properties, and how we employed it to generate networks for this study. For this study, we generated an ensemble of $500$ unique graphs using the Watts-Strogatz algorithm, and for each simulation, we sampled from this ensemble uniformly at random. 


Next, we assume that the initial perception of the common ground follows a Gaussian-like distribution, centred at $0$ certainty level, with standard deviation of $\frac{\sqrt11}{11}$. This represents a network of agents who are, at $t=0$, on average neutral about each piece of information belonging to the common ground. See the Supplementary Material for a description of how we realised this using a Beta distribution.
Then, for each $k\in \mathcal{I}$, we sample $x_{i,k}(0)$ from said distribution. Each simulation runs for a total of $T=m \times 5\times 10^5$. The number of time steps scales linearly with $m$, to compensate for the reduced selection probability of any given information as $m$ increases.  We selected this time window as preliminary simulations indicated this was an appropriate length of time to ensure convergence of the model dynamics. Due to the stochastic nature of \eqref{eq:sender_decision} and \eqref{eq:receiver_acceptance}, the state vectors $x_i(t)$ will not converge to constant vectors as $t\to\infty$, but for large $t$, one typically observes that the $x_i(t)$ will fluctuate around some static values. Future work may focus on theoretical analysis of the model, such as proving convergence of the expected value of $x_i(t)$.

As the focus of this paper is on the impact of different communication contexts, we assume homogeneity across a number of agent parameters to reduce the potential for confounding effects. In particular, we set the agent's susceptibility to social influence $\alpha_i = 0.2$ and their selection bias $\beta_i = 5$ for all $i\in \mathcal{V}$. This ensures that the rate of change of $x_{i,k}(t)$ is modest (i.e., each agent is only mildly susceptible to social influence for any given interaction), and agents are quite likely to select information they are certain belongs to the common ground.

\subsection{Identifying common ground formation}\label{ssec:metrics}

We now detail our approach to determining whether a common ground has formed with the network, either globally across all agents, or within a subgroup of agents.

Our methodology relies on identifying clusters of agents with ``similar'' state vector $x_i(t)$. This does not mean the agents have similar opinions or behaviours, but rather, they have a similar perspective of what information is known by other agents, and thus forms the common knowledge. This is because we have been deliberate in defining $x_i(t)$ to represent agent~$i$'s perception of what information other agents know of, they share what they believe to be known, and interpret the response. Agents who have similar state vectors are thus in agreement (consensus) about what information other agents know of, and thus part of the common ground, and communication grounding attempts between any such agents have a substantial chance of being successful.

Due to its robustness and its ability to deal with a variable number of clusters, we used Density-based spatial clustering of applications with noise (DBSCAN)~\cite{ester_1996_dbscan}. More details on the rationale for this choice and its implementation are reported in Appendix~\ref{app:dbscan}. Briefly, DBSCAN is able to identify clusters of points that are close to one another in Euclidean distance, and the number of detected clusters is not fixed a priori. Within a cluster, agents have sufficiently similar state vectors $x_i(t)$, and thus they are closely aligned in terms of how they perceive the common ground. As such, agents within a cluster are said to belong to the same common ground, and as will be apparent in the sequel, it is possible to have one cluster or multiple clusters (and the number of clusters can change over time). In other words, we can observe a global common ground (one cluster) or a fragmented common ground (multiple clusters). In the case of the latter, this corresponds to agents within a cluster sharing the same common ground, but the common ground of one cluster differs from that of another cluster. Those points which are not assigned to a cluster are termed ``noise points''. In context, if a data point is a noise point for our model, then the associated agent~$i$ is such that its perceived common ground is not aligned with a cluster. We interpret this by saying that agent~$i$ perception of the information environment is not reflective of the true state of the information environment, and what they believe belongs to the common ground is not consistent with the beliefs of other agents. That is, the agent does not share a common ground with any other agent.

\subsection{Emergence of global and local common ground}\label{ssec:example_sims}

To begin, we showcase exemplar simulations that illustrate the diverse emergent phenomena of the proposed SRG model, and discuss how to interpret these phenomena to gain insight into the process of common ground formation. We begin with a simple case of $m = 2$. That is, there are two pieces of information that agents can potentially share and contribute to the common ground. The decay factor is set to $\sigma = 0.05$. We consider two interaction contexts which represent generic offline ($\gamma = -1$, $\varepsilon = 0.1$) and online ($\gamma = 0$, $\varepsilon = 0.5$)  environments. See Section~\ref{ssec:motivation_model} above for the detailed justification.

\begin{figure}
    \centering
\subfloat[Time series of $x_i(t)$ in offline environment]{\includegraphics[width=0.7\linewidth]{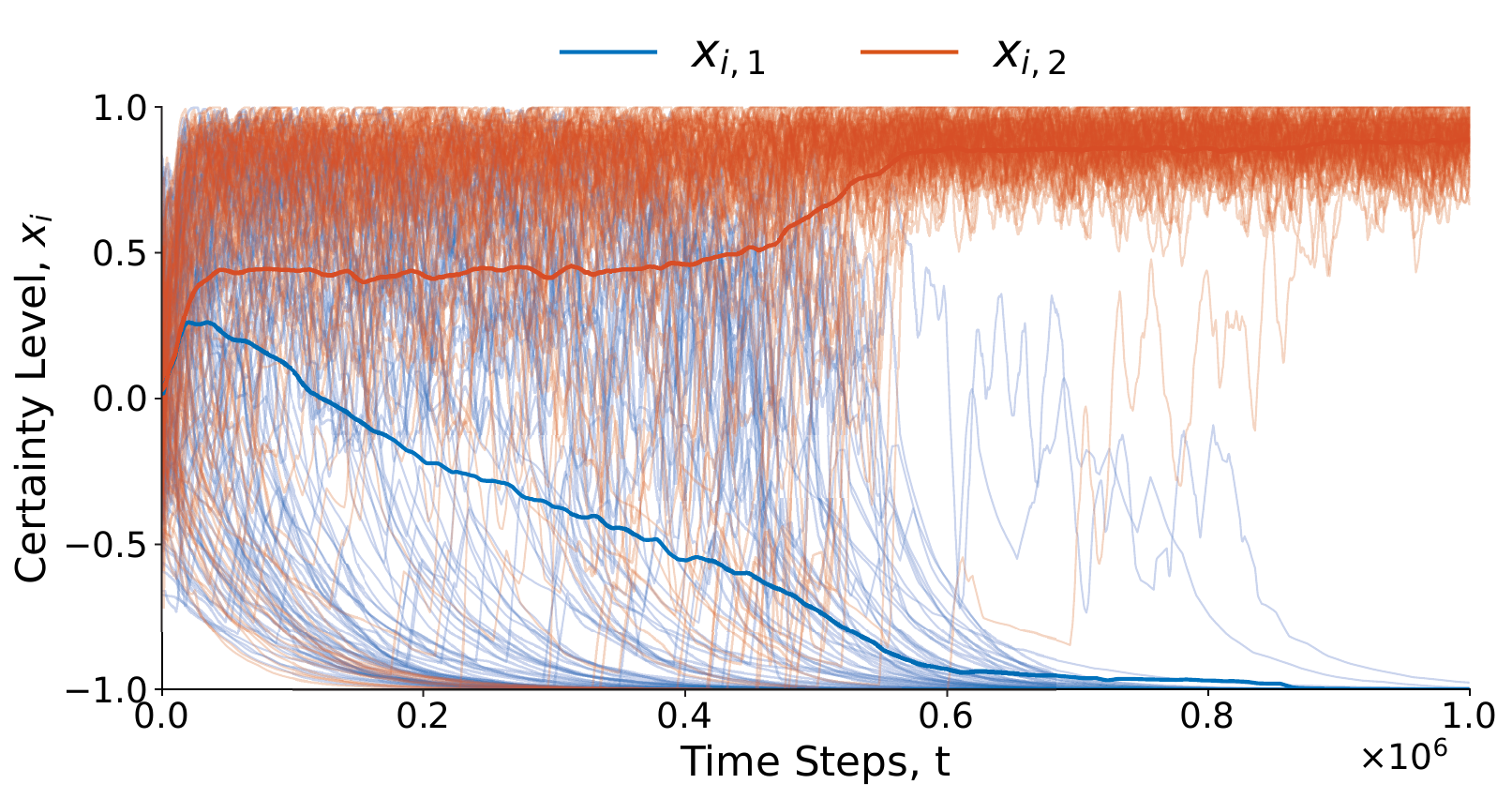}\label{fig:offline_TS}}
\vfill
\centering
\subfloat[Time series of $x_i(t)$ in online environment]{\includegraphics[width=0.7\linewidth]{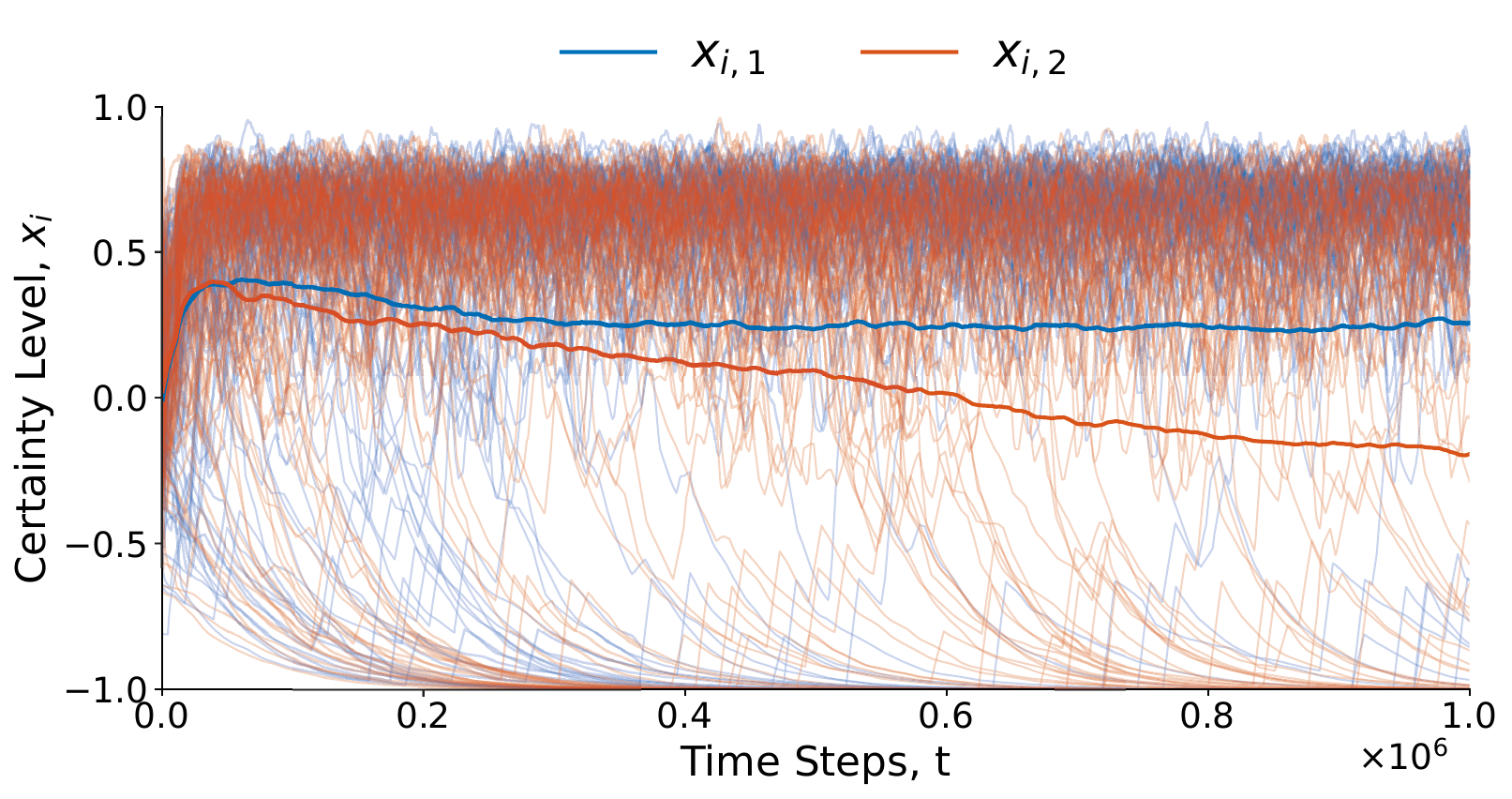}\label{fig:online_TS}}
    \caption{Time series illustrating example simulations of common ground formation in different information environments. Common parameters used are $n = 500$ agents 
    with $m = 2$ pieces of information. The time evolution of $x_{i,1}(t)$ (blue) and $x_{i,2}(t)$ (orange) is shown, thin lines being individual agents and thick line tracking the average $\frac{1}{n}\sum_{i=1}^n x_{i,k}(t)$. In (a) we have $\gamma = -1$ and $\varepsilon = 0.1$, and in (b) we set $\gamma = 0$ and $\varepsilon = 0.5$.}
    \label{fig:time_series}
\end{figure}

Figure~\ref{fig:time_series} shows two simulations; Fig.~\ref{fig:offline_TS} and Fig.~\ref{fig:online_TS} correspond to the offline and online information environment, respectively. In particular, we plot the time evolution of $x_i(t)$, with blue and orange indicating $x_{i,1}(t)$ and $x_{i,2}(t)$, respectively. Each thin line is a single agent, while the thick bold line tracks the average value, $\hat x_k(t) \triangleq \frac{1}{n}\sum_{i=1}^n x_{i,k}(t)$ for $k = 1,2$. Due to the large timescales and stochastic dynamics, we calculate the moving average of each agent's certainty level, taking a window of $10,000$ previous time steps. To enhance visibility, we plotted the certainty level every $500$ time steps, which is the average time window for an agent to have acted as a sender once.

Beginning with Fig.~\ref{fig:offline_TS}, we observe that $x_{i,1}(t) \to -1$ for all agents~$i\in \mathcal{V}$ as $t\to T = 1\times 10^6$. Meanwhile, information~$2$ survives and persists among all agents, with $\hat x_2(t) > 0.8$ as $t\to T = 1\times 10^6$. In context, at the beginning, both information are prevalent in the network and different agents have different certainty about the information that forms the common ground. However, through repeated iterations of the interaction and grounding process described above, agents are eventually maximally certain that information~$1$ is not part of the common ground, but highly certain information~$2$ is. By the end of the simulation, information~$1$ has died out (since agents almost always avoid transmitting it) and essentially only information~$2$ is communicated at each interaction. There is a global consensus on the common ground, i.e., a \textit{global common ground}. This is verified through DBSCAN, which identifies a single cluster of $500$ agents at time $T$. 

This is in stark contrast to the dynamics of the online information environment. In Fig.~\ref{fig:online_TS}, we observe that both information persist towards the end of the simulation. The average $\hat x_k(t)$ is less informative here, but we can clearly see that for both information~$1$ (blue) and information~$2$ (orange), agent certainties converge either to a high positive value (between $0.4$ and $0.8$) or a maximally negative value (around $-1$). While the time series plot allows us to determine that both information persist, it does not easily indicate the nature of the common ground, as we cannot easily distinguish whether for a given agent~$i$, $x_{i,1}$ and $x_{i,2}$ have the same sign. 


Here, application of DBSCAN identifies three distinct clusters. The first has $121$ agents, with an average state vector of $\bar x(T)=[-0.92, 0.67]$ and a total variance of $\text{Var}_{\text{total}}=3.47 \times 10^{-2}$ (computed as the trace of the covariance matrix). These agents are highly certain that information~$1$ is not part of the common ground and are certain that information~$2$ is. The second cluster has $135$ agents, with an average state vector of $\bar x(T)=[0.53, 0.43]$ and a total variance of $\text{Var}_{\text{total}}=1.11 \times 10^{-1}$. These agents are moderately certain that both information belong to the common ground. The third cluster has $243$ agents, with an average state vector of $\bar x(T)=[0.69, -0.96]$ and a total variance of $\text{Var}_{\text{total}}=2.01 \times 10^{-2}$. These agents are certain that information~$1$ is part of the common ground and are highly certain that information~$2$ is not. See the Supplementary Material for plots that visualize these clusters.


\begin{figure}
    \centering
\subfloat[Network snapshots in offline environment]{\includegraphics[width=\linewidth]{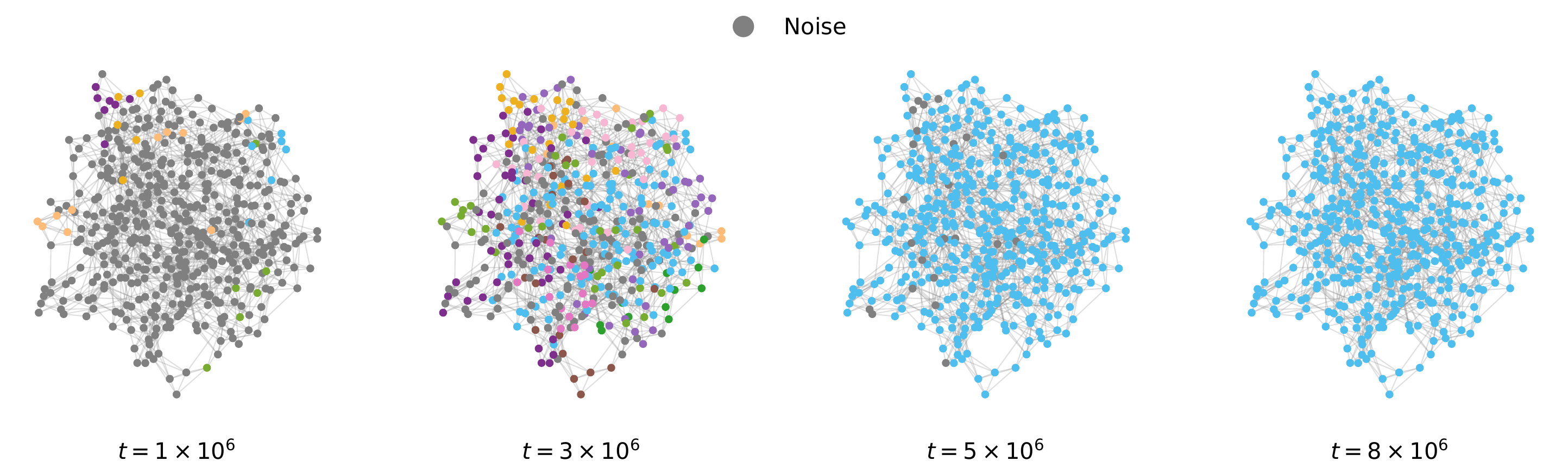}\label{fig:offline_network}}
\vfill
\centering
\subfloat[Network snapshots in online environment]{\includegraphics[width=\linewidth]{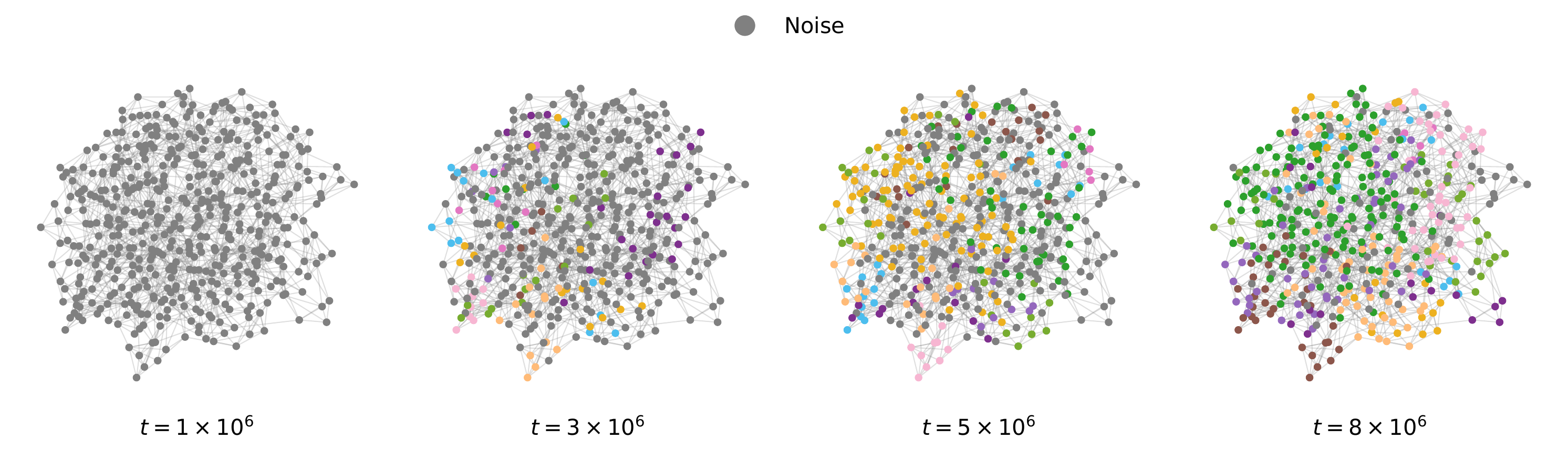}\label{fig:online_network}}
    \caption{Network snapshots of example simulations of common ground formation in different information environments. Common parameters used are $n = 500$ agents 
    with $m = 16$ pieces of information. Snapshots of the network at different time steps are shown, with different colours corresponding to different identified clusters. Grey nodes do not belong to any cluster, identified as noise using DBSCAN. In (a) we have offline environment ($\gamma = -1$, $\varepsilon = 0.1$), in (b) we have online environment ($\gamma = 0$, $\varepsilon = 0.5$).}
    \label{fig:network_example}
\end{figure}

We now showcase more expansive and complex phenomena, by increasing the number of pieces of information to $m=16$. Fig.~\ref{fig:offline_network} and \ref{fig:online_network} show the dynamics over the network for the offline and online environments, respectively, with each panel illustrating the state of the network at a point in time. For each panel, we ran DBSCAN to identify the clusters. Nodes belonging to the same cluster have the same colour, while grey nodes are those identified as noise points in DBSCAN. As discussed above, grey nodes are agents that do not belong to an identified cluster, and do not share a common ground with any other agent. In Fig.~\ref{fig:offline_network}, the offline environment, we see that in the initial period ($t=1\times 10^6$), there are a few small clusters, but mostly grey nodes. This suggests that a few small groups have reached a common ground, but most agents have different perceptions of the common ground. 
By time $t=3\times 10^6$, we observe that most agents have become part of local clusters as denoted by nodes coloured cyan, green, purple, yellow, etc. Note that the size (i.e., the number of members) of the clusters vary. Over time, a single cluster (cyan) emerges as the global common ground. A different process plays out in  Fig.~\ref{fig:online_network}, the online environment. In this case, we still observe a substantial number of nodes marked grey (i.e., the agents are not part of any common ground) even up to $t = 3\times 10^6$. At $t=5\times 10^6$, there are still a noticeable number of grey nodes, but also a diverse number of clusters, identified by the different coloured nodes. This diversity persists, and at $t = 8\times 10^6$, we see the grey nodes have largely disappeared but a range of different coloured clusters still exist. In other words, each agent now shares the same perception of common ground with another cluster of agents, and the different colours indicate that the perceived common ground differs across the clusters.

Comparing $m=2$ to $m=16$, we notice that in both settings, the offline environment results in a global common ground shared by all agents. Similarly, in the online environment, we see that agents are split into clusters whereby within a cluster, agents share the same perception of the common ground, and this perception differs between the clusters. However, a key difference is that the diversity is substantially greater for $m = 16$. In fact, DBSCAN reports $13$ clusters at $t = 8\times 10^6$ (Fig.~\ref{fig:online_network}), compared to the three clusters in Fig.~\ref{fig:online_TS}. Recall that the offline environment captures a low probability of signal loss ($\varepsilon = 0.1$) and the sender has a bias to interpret signal loss as a rejection ($\gamma = -1$). This evidently results in less information diversity (fewer clusters), and any common ground is likely to be adopted universally by all agents over time. In contrast, the online environment has a 50\% probability of signal loss, and any such event is interpreted as a neutral outcome by the sender. In such an environment, the overall common ground formation process is much slower, with clusters beginning to emerge only at $t = 3\times 10^6$, and not substantially forming until $t = 5\times 10^6$. 

In both Figs.~\ref{fig:time_series} and \ref{fig:network_example}, a key noticeable difference between the online and offline environment is whether there is a global common ground or not. Such an outcome can be related to the grounding process, and the impact of the two parameters $\varepsilon, \gamma$. In the offline environment, \eqref{eq:sender_update} indicates that the certainty of the sender $i$, $x_{i,k}(t)$, will decrease when there is a clear rejection or no response, and the latter event occurs with a low probability ($\varepsilon = 0.1$). This means the sender and receiver almost always have the same interpretation of the grounding process outcome, and people are able to rapidly establish which information is part of the shared common ground.
In contrast, in the online environment, the sender fails to receive a signal  with a 50\% probability. Thus, a misalignment in how the sender and receiver interprets the grounding process outcome occurs more frequently. Importantly, the interpretation bias $\gamma = 0$ means the certainty $x_{i,k}(t)$ moves towards $0$ whenever this misalignment occurs, allowing more information to survive for longer. Over time, this helps to facilitate the emergence of local common ground clusters.

We have now presented illustrative example simulations of the SRG model, for different interaction contexts (captured by $\gamma$ and $\varepsilon$) and when there are few or many pieces of information to select from (captured by $m$). The simulations suggest both global and local common ground can form, with $\gamma$, $\varepsilon$ and $m$ influencing the number of different common ground that form over time, as well as the speed of the formation process. In the next chapter, we present our main results, using a campaign of Monte Carlo simulations to provide a comprehensive and robust account of how $\gamma$, $\varepsilon$ and $m$ impact that common ground formation process.

\section{Main Results}\label{sec:results}

Having presented the SRG model and illustrated the variety of phenomena that emerges using several example simulations, we now turn to our main results. Namely, we explore how the number of pieces of information, $m$, and the interaction context (captured via $\gamma$ and $\varepsilon$) shapes the common ground formation process, both over time and in terms of the diversity of common ground that may emerge.

\subsection{Information diversity and survival over time}

\begin{figure}
    \centering
   \subfloat[$m=2$]{\includegraphics[width=0.48\linewidth]{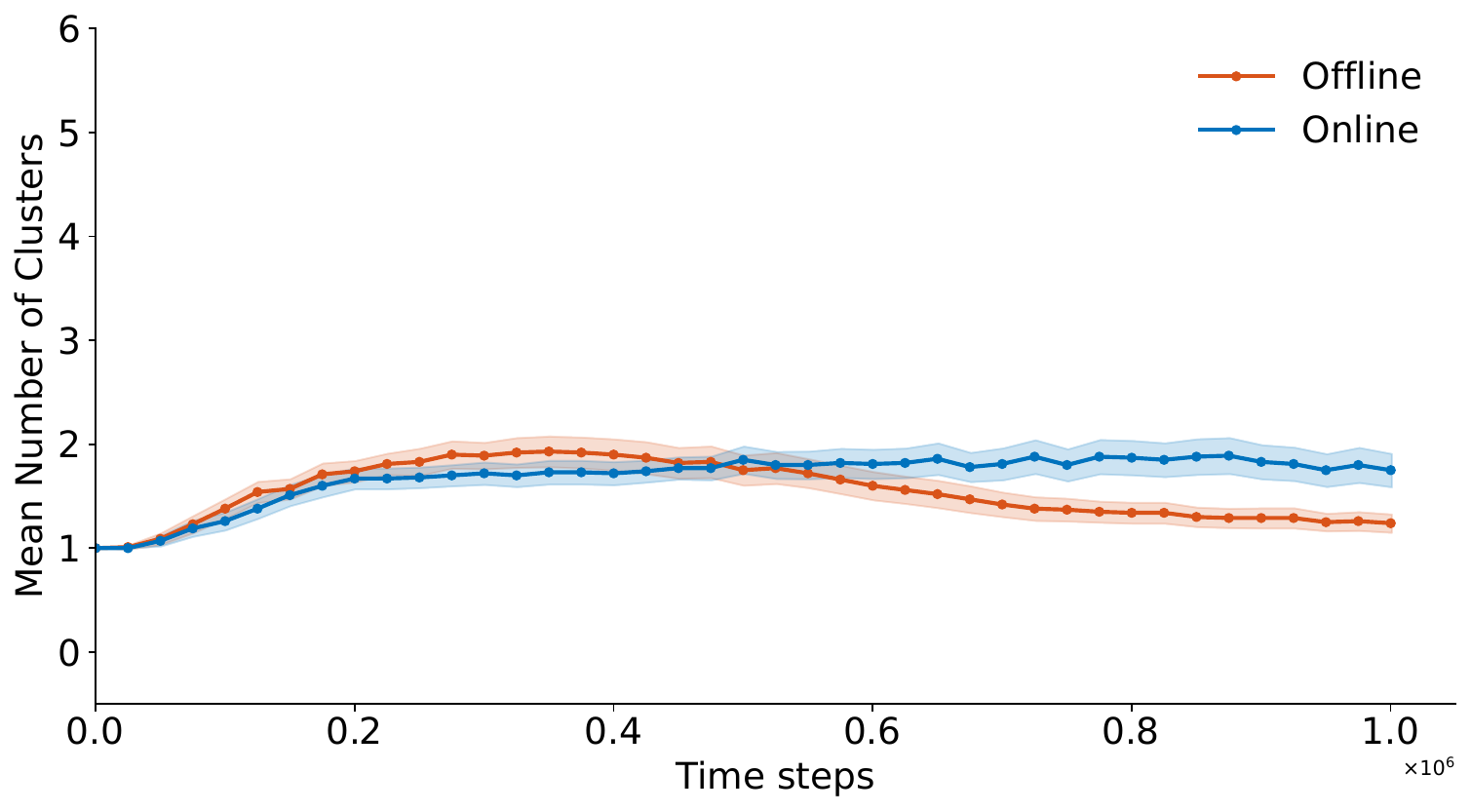}\label{fig:cluster_timeseries_m2}}
    \hfill
    \subfloat[$m=4$]{\includegraphics[width=0.48\linewidth]{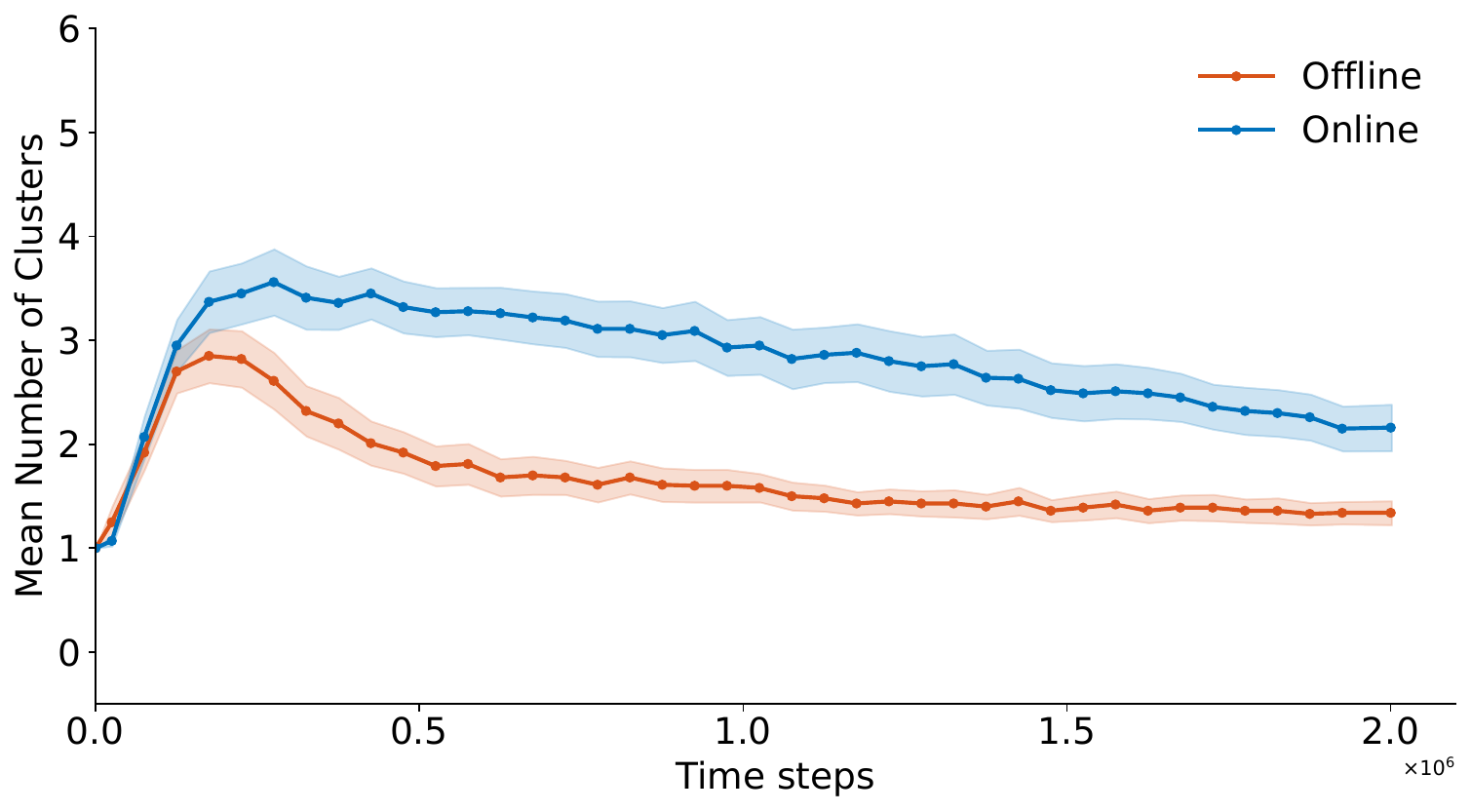}        \label{fig:cluster_timeseries_m4}}
\vfill
   \subfloat[$m=8$]{\includegraphics[width=0.48\linewidth]{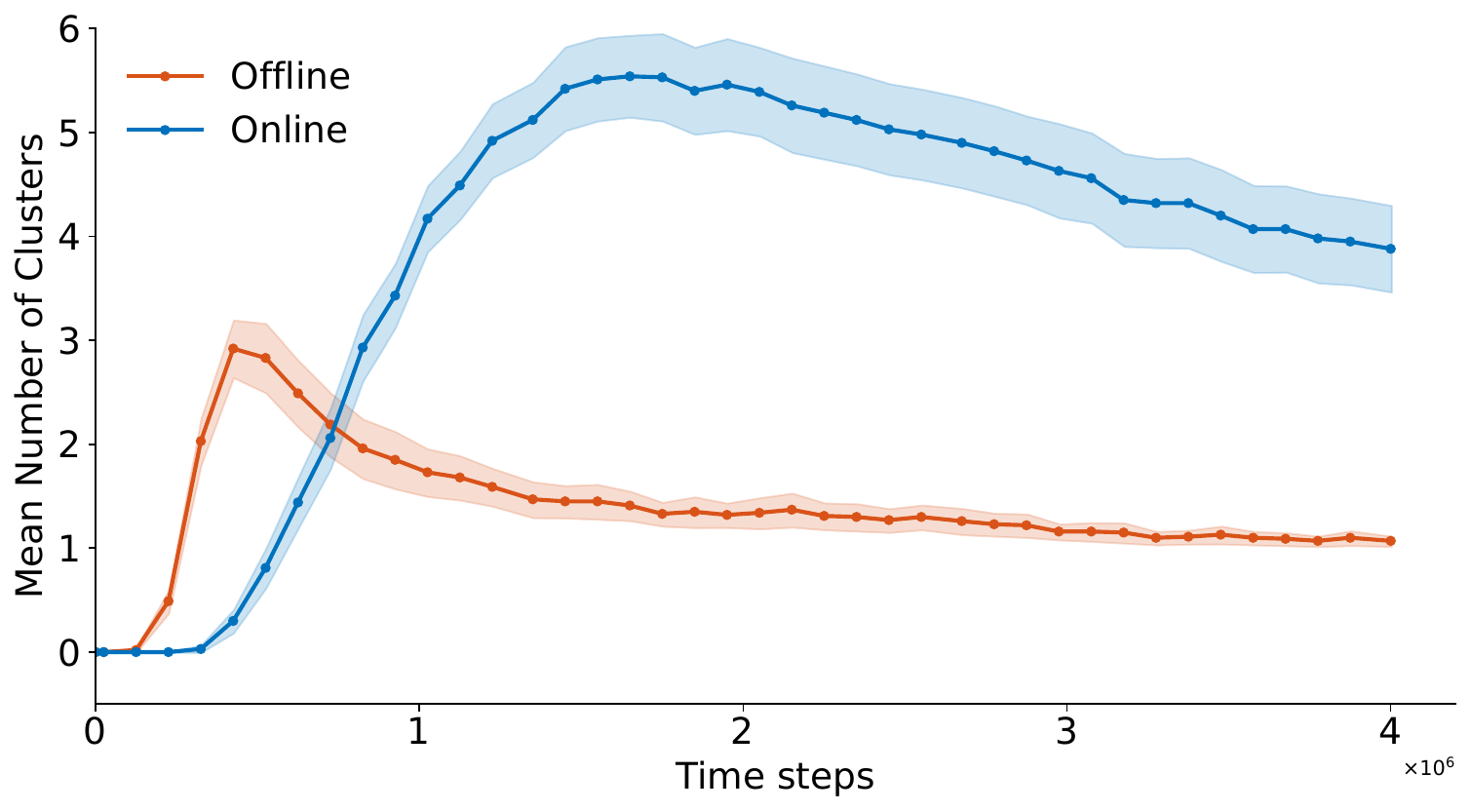}
        \label{fig:cluster_timeseries_m8}}
    \hfill
    \subfloat[$m=16$]{\includegraphics[width=0.48\linewidth]{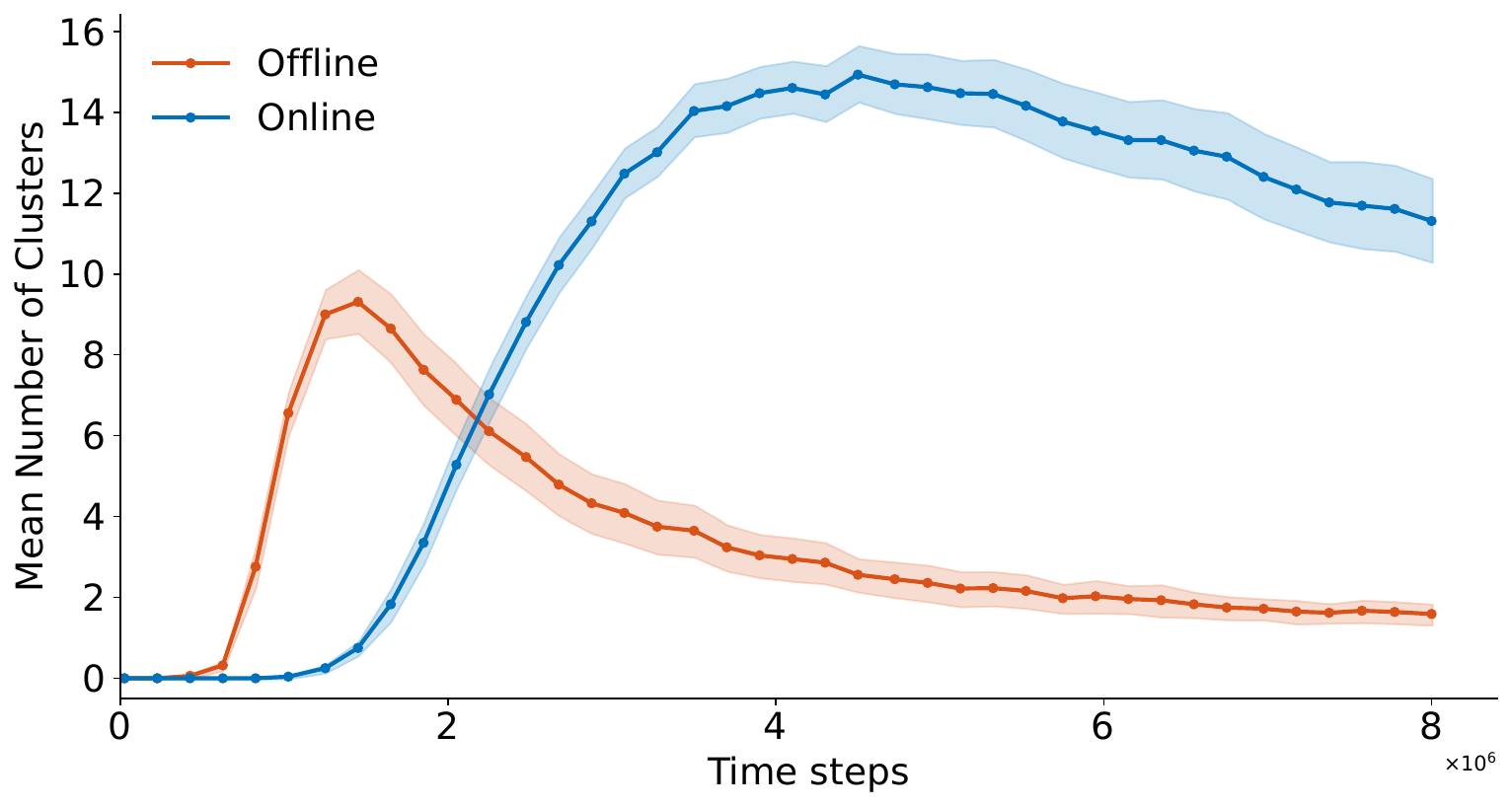}\label{fig:cluster_timeseries_m16}}
    \caption{Time series of the number of clusters for offline ($\gamma = -1$ and $\varepsilon = 0.1$) and online ($\gamma = 0$ and $\varepsilon = 0.5$) settings at various values of $m$, i.e., the number of information available to be communicated. The solid line indicates the mean of 100 Monte Carlo simulations, with the shaded region being a 95\% confidence interval. Note panel (d) has a difference y-axis scale. }
    \label{fig:cluster_timeseries}
\end{figure}

To begin, we consider a continuation and extension of the simulations presented above. We use the simulation setup in Section~\ref{ssec:sim_setup}, and for each of $m = 2, 4, 8, 16$, we simulate the two scenarios previously considered: (a)  $\gamma = -1$ and $\varepsilon = 0.1$, capturing an offline-type interaction context and (b) $\gamma = 0$ and $\varepsilon = 0.5$, capturing an online-type interaction context. For each setting, we run Monte Carlo simulations, with $100$ replicates. Figure~\ref{fig:cluster_timeseries} plots the number of clusters (identified using DBSCAN as detailed in Section~\ref{ssec:metrics}) over time, with the solid line capturing the mean over $100$ replicates, along with a 95\% confidence interval envelope.

First, we observe that at the end of the simulation time window, there are more clusters in the online setting ($\gamma = 0, \varepsilon = 0.5$) than in the offline setting ($\gamma = -1, \varepsilon = 0.1$). This property holds regardless of $m$, although the difference increases as $m$ increases. In other words, online interaction consistently produces more diversity in the long run, with multiple clusters of agents that share the same communal common ground, but the common ground differs between clusters. In the offline setting, there is typically a global common ground at the end of the simulation, except in $m=16$, where the average is two clusters. This is consistent with our exploratory simulation in Fig.~\ref{fig:network_example}, and in the next subsection, we explore this long-run diversity in more detail.

For the moment, we turn our attention to the transient dynamics, i.e., the evolution of the clusters over time. Interestingly, the diversity of the information (captured by way of proxy by counting the number of clusters) initially \textit{increases} to a peak, before decreasing over time. We conjecture that this is a reflection of the grounding process as occurring in our model. Early in the simulation, the  perceived common ground of the agents are not aligned and the average certainty is close to neutral, $x_{i,k}(t) = 0$. This means agents will communicate a broad range of information to each other initially, and the grounding process allows agents to learn what belongs to the common ground. The highly clustered nature of the Watts--Strogatz network facilitates the initial formation of a greater number of clusters, and the number decreases as clusters start to merge when agents interact across clusters. 

The transient dynamics are much shorter for the offline environment, in the sense that the number of clusters settles much more quickly, although somewhat curiously, the temporary increase at the start of the simulation is also faster in offline environments. This suggests that not only is there typically a global common ground formed in the offline environment ($m = 2,4,8$), but the formation rate is much faster than the online environment. We believe the parameter $\varepsilon$ plays a pivotal role here. With a small $\varepsilon$, sender agents in the offline environment almost always observe the response signal, and the sender is able to increase or decrease their certainty $x_{i,k}(t)$ on the information $k$ they transmitted, regardless of whether the receiver rejects or accepts it. In contrast, in the online environment, senders have a 50\% chance of not observing the response signal, and because $\gamma = 0$, their certainty moves towards a neutral point. This results in a much longer period for establishing common ground.

Finally, we note that as $m$ increases, the number of clusters in the online environment increases dramatically from 2 clusters ($m=2$) to 12 clusters ($m=16$). Intuitively, the more options there are to initially select from (large $m$), the more diverse the common ground is in the long run: although agents are aiming to reach a consensus on what they perceive to be the common ground, a large number of options can result in local consensus but global diversity, especially due to the clustered structure of the Watts--Strogatz network. Such an outcome accords with several other agent-based models of cultural dynamics and social norm formation~\cite{peytonyoung2015social_norms,Axelrod1997}. Evidently, online-type interactions are more conducive to maintaining this diversity (as already noted above). In the next subsection, we explore how $m$, $\gamma$, and $\varepsilon$ shape the long term diversity of the common ground.

\subsection{Interaction context effects on long-term information diversity}

\begin{figure}
    \centering
   \subfloat[$m=2$]{\includegraphics[width=0.48\linewidth]{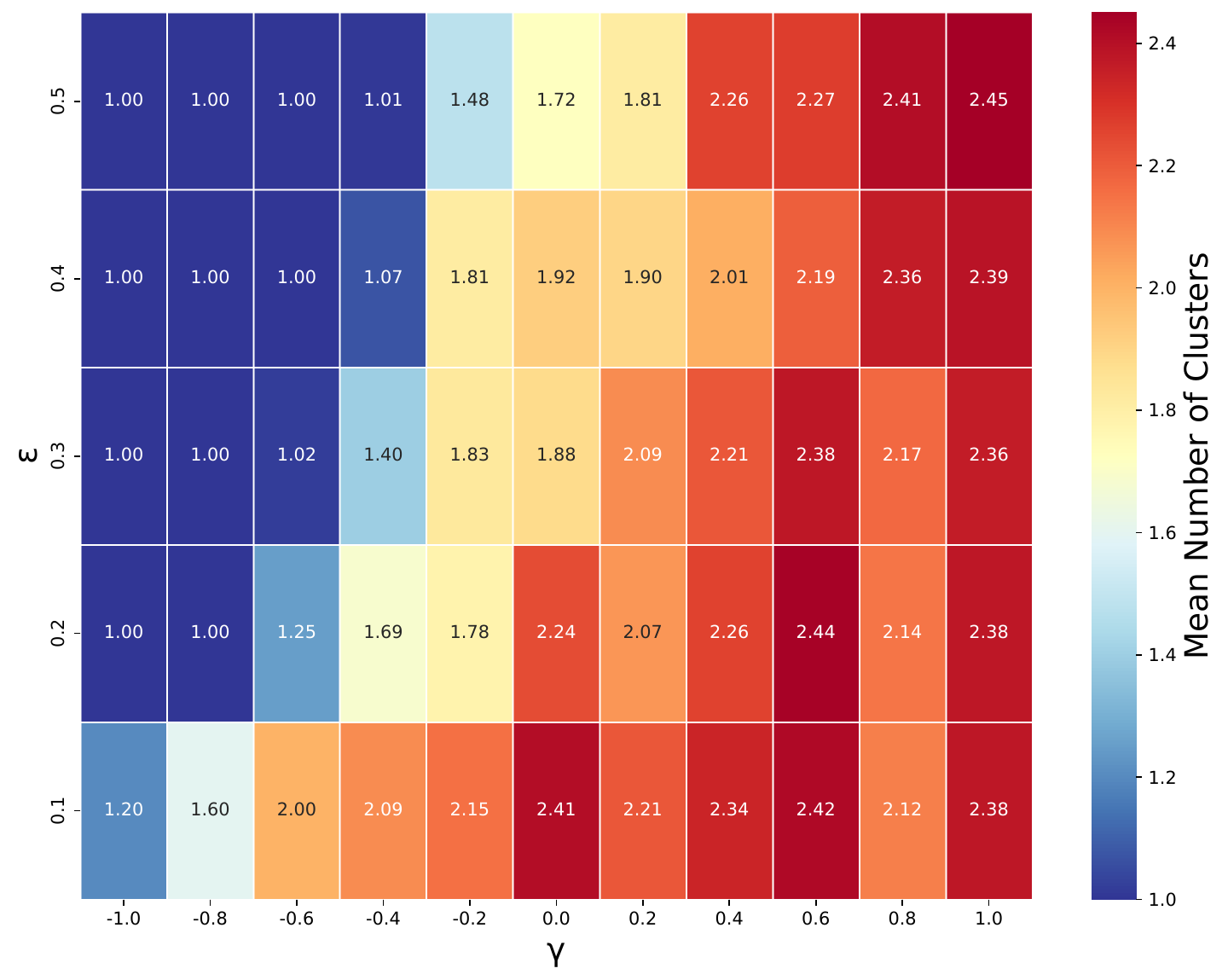}\label{fig:cluster_heatmap_v2_m2}}
    \hfill
    \subfloat[$m=4$]{\includegraphics[width=0.48\linewidth]{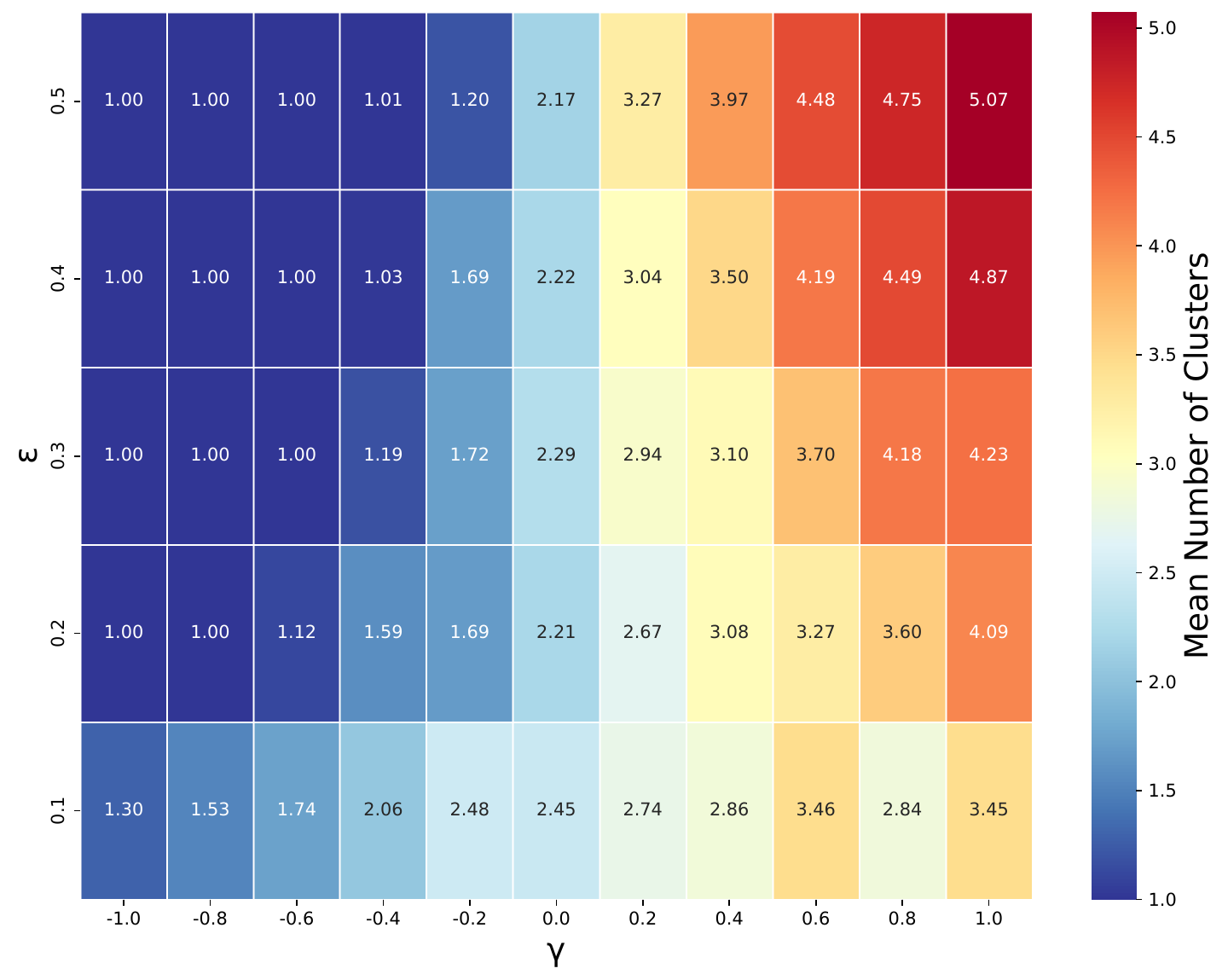}        \label{fig:cluster_heatmap_v2_m4}}
\vfill
   \subfloat[$m=8$]{\includegraphics[width=0.48\linewidth]{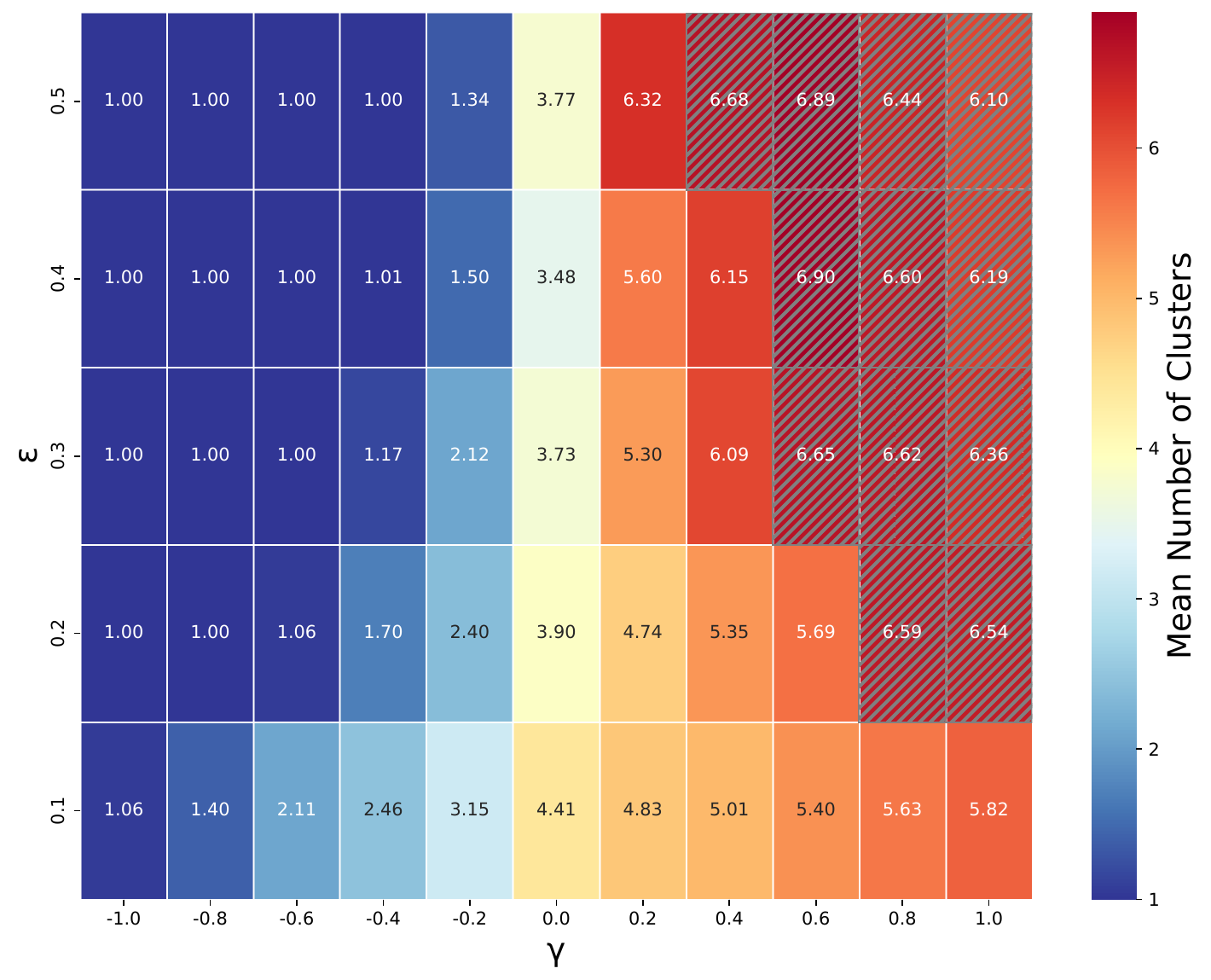}
        \label{fig:cluster_heatmap_v2_m8}}
    \hfill
    \subfloat[$m=16$]{\includegraphics[width=0.48\linewidth]{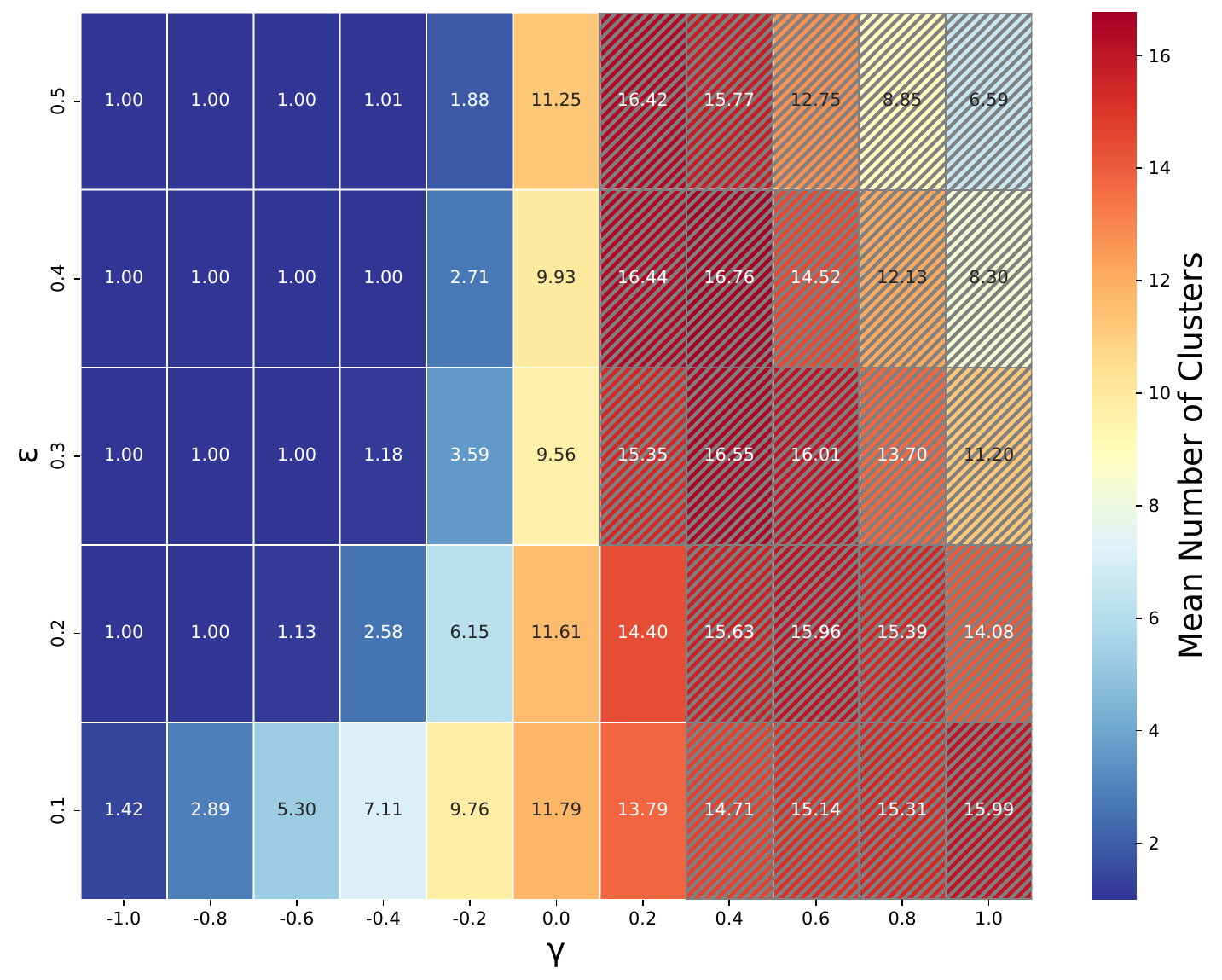}\label{fig:cluster_heatmap_v2_m16}}
    \caption{Heatmaps of the number of clusters as a function of $\gamma$ and $\varepsilon$, at various values of $m$. The colour intensity indicates the mean number of clusters at the end of the simulation time, averaged over 100 Monte Carlo simulations. Greyed out cells correspond to simulations in which DBSCAN reported over $30\%$ of the data points (nodes) as being noise.}
    \label{fig:cluster_heatmap}
\end{figure}

Having explored the evolution and formation of common ground clusters over time in Fig.~\ref{fig:cluster_timeseries}, we now turn our attention to examine the long-term information diversity and final common ground clusters (i.e., at the end of the simulation time window), as a function of the interaction context. In particular, we vary interpretation bias $\gamma$ from $-1$ to $1$ at steps of $0.2$, and we vary the probability of signal loss $\varepsilon$ from $0.1$ to $0.5$ at steps of $0.1$. For each pair $\gamma, \varepsilon$, we run Monte Carlo simulations with $100$ replicates. The mean number of clusters at the end of the simulation time window are plotted as heatmaps in Fig.~\ref{fig:cluster_heatmap}, for different numbers of pieces of information, $m = 2, 4, 8, 16$. Note that each panel has a different scale in terms of colour intensity, as the number of surviving clusters vary substantially for different $m$. Some cells are greyed out. These correspond to simulations in which DBSCAN reported over $30\%$ of the data points (nodes) as being noise. Such a high level of noise corresponds to a scenario in which there is no shared common ground, which we elaborate on below. In the Supplementary Material, we explain the issue of noise in DBSCAN, how we determined which cells in Fig.~\ref{fig:cluster_heatmap} contain too much noise, and the justification for interpreting high levels of noise as the lack of a shared common ground.

The first observation we wish to point out is that for any $m$, a global common ground is a possible emergent phenomenon and occurs for a range of $\varepsilon$ and $\gamma$ values (see Fig.~\ref{fig:offline_network} for an illustration of the formation of a global common ground). The likelihood of a global common ground emerging increases as $\varepsilon \to 0.5$ and $\gamma \to -1$, and in the deep blue cells, it is the only emergent population-level phenomenon. In other words, as the probability of signal loss increases, and as the sender increasingly interprets the absence of a response as an outright rejection, a global common ground will emerge. 
Recall that the grounding process is used by agents to determine which information belongs to the common ground. 
The amount of rejections (both actual and perceived by the sender when there is no response) increases as $\varepsilon \to 0.5$ and $\gamma \to -1$, and this means many pieces of information are rapidly removed from consideration for belonging to the common ground. Importantly, this occurs irrespective of $m$, suggesting that regardless of the diversity of information choice,  interaction contexts which result in explicit rejection or acceptance feedback to the sender facilitates a global communal common ground. 

Our second result is to observe that, when the interpretation bias $\gamma$ is non-positive, there is increased clustering and the occurrence of multiple local common ground for small $\varepsilon$. This provides additional nuance to the commentary in the prior paragraph. Namely, there is a subtle but important distinction between the sender observing the true outcome from the receiver (this occurs with high probability when $\varepsilon$ is small) and the sender not observing the true outcome but rather a perceived rejection (occurring when $\varepsilon \to 0.5$ and $\gamma \ll 0$). In the case of the former, there is no discrepancy between the sender and the receiver when updating using \eqref{eq:agent_update}, whereas there is substantial discrepancy in the latter with the sender more likely to perceive rejection even if the receiver actually accepted the transmission. This discrepancy may result in more information being removed from consideration as the sender is more likely to consider transmitted information as not belonging to the common ground due to the greater amount of perceived rejections. In contrast, a closer alignment between sender and receiver in terms of outcome of the grounding process means that even when $\gamma$ is negative, multiple information can survive and one can have multiple local common ground (clusters).

Finally, we point out that positive interpretation bias $\gamma$ dramatically increases the diversity of the information surviving in the long term. For small numbers of options $m=2,4$, the emergent global phenomenon is the existence of multiple clusters, where agents within a cluster share the same common ground, and the common ground between clusters are substantially different. Fig.~\ref{fig:online_network} shows an illustrative example of the multiple clusters. On the other hand, when $m = 8, 16$ (there are many options to select) we observe that as $\gamma$ increases (and also as $\varepsilon$ increases), there is a total collapse of shared common ground as indicated by the greyed out cells in Fig.~\ref{fig:cluster_heatmap_v2_m8} and \ref{fig:cluster_heatmap_v2_m16}. These greyed out cells represent when DBSCAN associates over $30\%$ of the nodes as being noise and do not belong to any cluster. In the context of the model, there is not a shared common ground among the agents. Specifically, since $x_i(t)$ captures agent~$i$'s certainty of information belonging to the common ground, the fact that the $x_i$'s are not in a cluster can be interpreted as that agent~$i$'s perception of the common ground is inconsistent with that of other agents, and there is in fact no common ground of any sort. Agents live in their own information bubble, because with $\gamma > 0$, the sender interprets the lack of a response from the receiver as a successful grounding attempt, so the sender continues to believe the information is part of the common ground. This lack of negative feedback, along with an increased number of options to select from, means that stable clusters cannot form in the long term.


In the Supplementary Material, we provide additional information on the simulations to highlight the robustness of our findings. First, we show that the intra-cluster distance is at least an order of magnitude smaller than the inter-cluster distance, indicating that the common ground between different clusters are indeed substantially distinct. Second, we analyse the subnetworks arising from the DBSCAN algorithm, and find that almost always, the clusters identified in the DBSCAN algorithm correspond to a large connected component in the network. This illustrates that the agents which share a common ground (have similar $x_i$) are in fact connected to each other, and this common ground is a genuinely shared social construct arising from repeated interactions over the social network, rather than some artifact of the model dynamics.

\section{Discussion and Conclusion}\label{sec:discussion}

In this paper, we proposed the novel Sender--Receiver Grounding (SRG) model, taking an agent-based approach and focusing on the micro-level cognitive process of \textit{grounding} between a sender and a receiver. Our simulations focused on how three key factors in the model contributed to shaping the emergent macro-level phenomena ---the number of pieces of information available to select from, the likelihood of a response signal not being observed by the sender, and how the sender interprets this lack of response. 

The model predicted a range of different macro-level phenomena, including the formation of a global communal common ground, the emergence of multiple clusters capturing a fragmented communal common ground, and the state of \textit{anomie} in which agents live in their own information bubble. How the sender interprets the lack of a response (the parameter $\gamma$ in the model) plays a central role in shaping said emergent phenomena. 

When $\gamma < 0$, the sender interprets the lack of a response as a rejection, and a global communal common ground is likely to emerge regardless of the number of available information to select from, $m$. The only caveat is when there is only a small likelihood of receiving no response (i.e., $\epsilon \approx 0$). In this case, fragmentation of the common ground is observed, with the number of clusters increasing as $m$ increases. This is an important and nuanced point: clear and unambiguous grounding interactions are not sufficient for global communal common ground. There needs to be a response from a receiver in some form. Moreover, notice that  when a global common ground, it does so over a long period of time and multiple interactions between agents. During the initial transient period, the simulations indicate that many clusters form initially, and then merge together over time. As $\gamma$ approaches $0$, the sender adopts a neutral/uncertain interpretation of the lack of a response, and we consistently observe a fragmented communal common ground. Interestingly, the number of clusters is primarily impacted by the available of pieces of information $m$, while $\epsilon$ plays only a small role. Finally, once $\gamma$ increases above $0$, the available information $m$ controls the emergent macro-level outcome. For small $m=2,4$, fragmentation occurs, with agents clustering into different communal common ground. For larger $m = 8,16$, a complete atomisation occurs, resulting in a state of \textit{anomie}, with each agent living in their own information bubble.

There are several implications from these insights. First, and as we highlighted in the Introduction, establishing a global communal common ground is critical for collective action, e.g., on climate action or pandemic response. Besides top-down broadcasting, traditional approaches to achieving consensus include individual-level interventions or introducing opinion leaders~\cite{ye2021collective,centola2018experimental_tipping,zino2022dynamic_norms}. Here, our findings suggest another approach would be to ensure agents interact in an environment with $\gamma \approx -1$; this could be achieved by establishing norms about communication which regard the lack of a response as a rejection, or through deliberate design of online social media platforms and how users interact with one another. Conversely, to prevent fragmentation of common ground or a state of anomie, one should aim to reduce the number of available information to select from as well as to avoid developing communication protocols that allow for the lack of a response to be interpreted as acceptance. This last point highlights the challenges in preserving and protecting the online information environment~\cite{noauthor_online_2022}, where many online platform designs are such that the sender is often not notified of responses, and where there are many alternatives and choices (including those deliberately injected into the environment during disinformation campaigns~\cite{paul2016russian_firehose}).

We conclude by discussing various directions for future research. In terms of extending the SRG model, three possibilities come to mind. First, we have considered in this paper pairwise asynchronous interactions. In many communication environments, one might have a single sender with multiple simultaneous receivers, such as an opinion leader or influencer making a statement. It may be interesting to examine how this impacts the common ground formation, and whether different common ground emerge around highly central nodes in the network. Second, in \eqref{eq:receiver_update}, we assume no change occurs for the receiver under rejection. This could be revisited by changing \eqref{eq:receiver_update} so that the receiver's certainty might decrease or increase with rejection (for instance if the receiver has an adversarial or friendly relationship with the sender, respectively). Third, the literature on common ground and the grounding process highlights that in different social contexts, a person may select different information to transmit. For instance, a sender who wishes to strengthen their social connection with the receiver will choose to send common ground consistent information to maximise the likelihood of acceptance, whereas a sender wishing to showcase their knowledge may send common ground inconsistent information. Our current model is closer to the former, and thus it would be of interest to examine how the dynamics change if we implement a mechanism more reflective of the latter.

In terms of applying the model to study relevant common ground formation problems, three possibilities directly flow from our work. First, we have only considered the Watts-Strogatz network model, which has small-world properties (short characteristic path length, high clustering) like many real-world networks, but does not have hub nodes or a power-law degree distribution. Other network models, and real-world network examples, can be used in order to understand the impact of different network topological characteristics on common ground formation. Second, the ``survival-of-the-fittest'' problem is a classical one in cultural and evolutionary dynamic. In this paper, all information are equivalent. It would be interesting to embed biases so that some pieces of information are ``fitter'' than others (adopting the terminology of evolutionary biology), and examine the likelihood of the fitter information surviving, modulated by the network topology and other agent-level parameters and mechanisms. A standard approach, often used in evolutionary game theory, is to consider a population initially sharing global common ground, then injecting new information in a small fraction of the population, before observing whether the new information can successfully invade and become endemic in the population. Finally, information diffusion is another classical problem that should be examined using this model. It would be of interest to replicate existing studies, which insert source agents (who persistently transmit a single type of information), and examine the speed and depth of the diffusion process.


\section*{Appendix}

\subsection{Watts--Strogatz network}\label{app:ws}

In this paper, all networks are generated according to the Watts--Strogatz algorithm (as implemented in Python's NetworkX package), which is a stochastic algorithm that generates small-world networks. In particular, the network is generated as follows. First, we construct a regular ring lattice with $n$ nodes, where each node is connected to the $k$ nearest neighbours ($k/2$ on each side, and thus it is typically assumed that $k$ is even). Then, we fix a rewiring probability $p$. Each edge of this lattice is rewired with probability equal to $p$, independently of the other edges. Edge rewiring is performed by removing the edge from the edge set and substituting it  with a new edge that keeps one fixed endpoint and connects it to another node chosen uniformly at random among the remaining nodes that are not already connected to it. In our simulations, we have set $k = 6$ and $p =0.2$

According to this procedure, the network obtained has several important features of real-world networks. In particular, the network is characterized by a large clustering coefficients (i.e., high rate of triadic closures, whereby a large fraction of the neighbours of neighbours tend to be connected) and a small diameter, which are both features of many real-world social networks~\cite{newman2010networks_book}. 

For this study, we generated an ensemble of $500$ unique graphs using the Watts-Strogatz algorithm with $k=6$ and $p = 0.2$, and for each simulation, we sampled from this ensemble uniformly at random.

\subsection{Details on DBSCAN}\label{app:dbscan}

After testing a number of different clustering methods (partition-based, hierarchical, density-based, network-based), we found that a method for reliably identifying clusters is Density-based spatial clustering of applications with noise (DBSCAN)~\cite{ester_1996_dbscan}. This is because the number of clusters can vary substantially both over time in a single simulation, and also with different model parameters. Thus, partition-based methods, which typically require a priori setting of the number of clusters, were not suitable. Moreover, given the highly stochastic nature of our model, hierarchical clustering methods were also not suitable. Network-based methods were also difficult to employ, since these typically identified clusters based on network topological characteristics, whereas we are interested in clustering based on the agents' time-evolving states. Among the various density-based methods, DBSCAN is a known and robust algorithm.

DBSCAN involves two parameters, $R \in \mathbb R_+$ and $MinPts \in \mathbb N_+$. Here, $R$ defines the maximum distance between two points to be considered neighbours and thus part of a cluster. In our case, each data point is associated with one agent, and in particular, the state vector $x_i(t)$ as plotted into $\mathbb R^m$. We use the Euclidean distance as the distance considered in the algorithm. Meanwhile, $MinPts$ is the minimum number of data points within a distance of $R$ to a specific point, for that specific point to be considered a core point. These core points are the starting points of a cluster, and a cluster is a set of points in which each point is within a distance of $R$ from at least one other core point in the cluster. After substantial experimentation, we determined that a suitable parameter setting was $R = 0.35$, and $MinPts = 26, 22, 14, 5$ for $m = 2, 4, 8, 16$, respectively.

 \bibliographystyle{IEEEtran}
 \bibliography{MYE_ANU}

\title{\LARGE \textsc{Supplementary Material}: An agent-based model of the formation and evolution of common ground}

\setcounter{figure}{0}
\renewcommand\thefigure{S\arabic{figure}}
\setcounter{table}{0}
\renewcommand\thetable{S\arabic{table}}
\numberwithin{equation}{section}
\renewcommand\theequation{\thesection\arabic{equation}}
\renewcommand{\thesection}{S\arabic{section}}
\setcounter{section}{0}



\begin{center}
    \LARGE \textsc{Supplementary Material}: An agent-based model of the formation and evolution of common ground
\end{center}


\section{Initial condition distribution}

\begin{figure}[htbp]
    \centering
    \includegraphics[width=\linewidth]{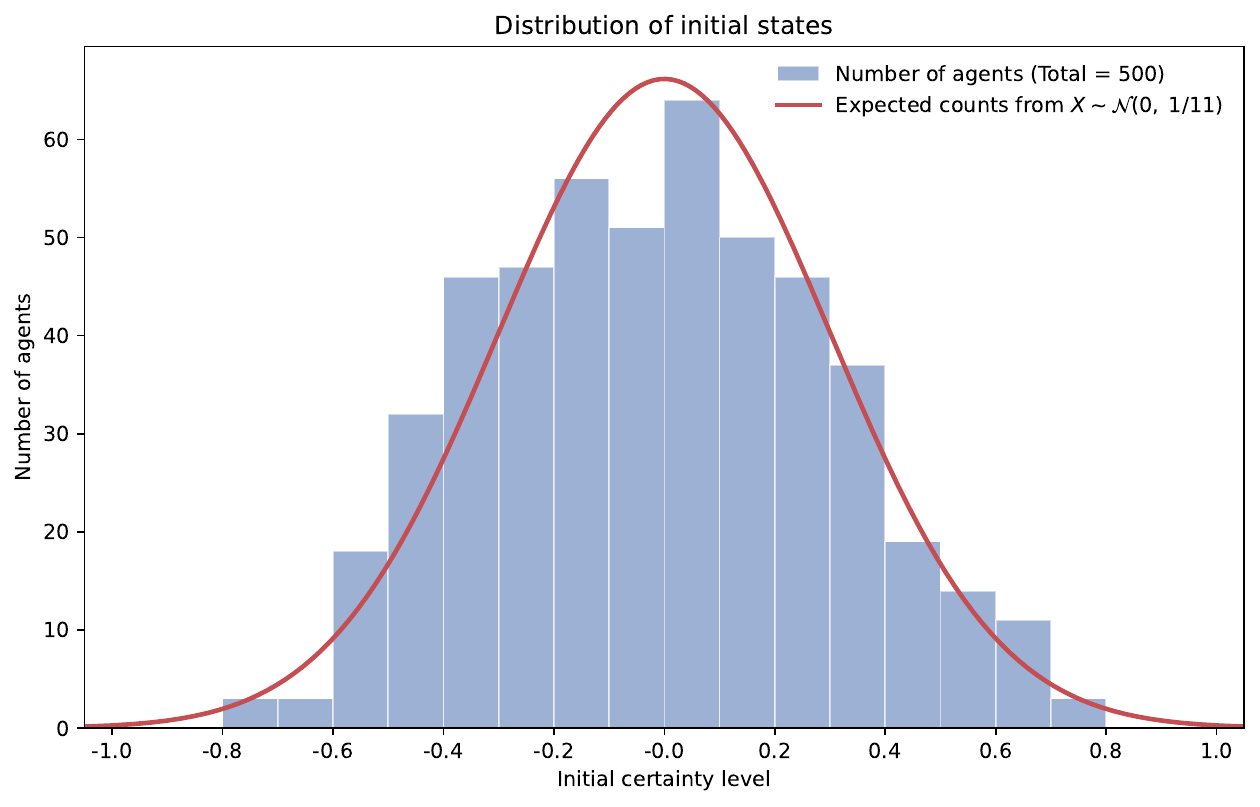}
    \caption{A typical distribution of initial agent certainty level, $x_{i,k}(0)$, for one piece of information $k$. The red line traces the expected number of agents if the certainty level were normally distributed with a mean of $0$ and a variance of $\frac{1}{11}$, which correspond to the mean and the variance values of the distribution we used ($2Beta(5,5)-1)$.}
    \label{fig:Initialisation_histogram}
\end{figure}

For each $k = 1,\hdots, m$, we sampled $x_{i,k}(0)$ from a probability distribution that was created as follows. We first generated a $Beta(a,b)$ distribution, with $a=b=5$, and then rescaled from $[0,1]$ to $[-1,1]$ using the mapping $x \mapsto 2x - 1$. Each $k$ was generated independently, for each simulation replication. Fig.~\ref{fig:Initialisation_histogram} shows an example of the distribution of initial certainty levels, $x_{i,k}(0)$, for one piece of information~$k$.

\section{Identifying Common Ground}

\subsection{DBSCAN}

Our strategy to identify common grounds is by grouping agents based on their perceptions of the common ground, $x_{i}\in m$. To this end, we treat each $x_{i}$ as a data point in a $[-1,1]^{m}$ hypercube, then employ a popular clustering algorithm known as \textbf{DBSCAN}, or Density-Based Spatial Clustering of Applications with Noise. DBSCAN identifies clusters based on the pairwise similarity between the vectors. In our implementation, we elect to use the Euclidean distance, which is a common choice. 

Most popular clustering algorithms can be classified as one of: 1. \textbf{Partitioning clustering} 2. \textbf{Hierarchical clustering} or 3. \textbf{Density-based clustering}. Partitioning clustering methods, such as $k$-means clustering, require the number of clusters to be pre-defined. The stochastic nature of our model and the diverse emergent phenomena makes selecting $k$ a significant challenge. Additionally, partitioning clustering methods inherently assume ball-shaped clusters and there is no guarantee that agents with common grounds form ball-shaped neighbourhoods in Euclidean space. Hierarchical clustering methods, on the other hand, struggle with outliers or noise points because every point must be assigned to a cluster. This means that points in sparse regions in between clusters can end up causing two otherwise distinct clusters to be merged and counted as one cluster. This may be problematic because ``inbetweener" agents are often observed in our model which would act as the points in the sparse regions between clusters. Conversely, Density-based clustering methods, such as DBSCAN, do not require the user to pre-specify the number of clusters, do not make assumptions about the shape of clusters, and is able to classify noise points. DBSCAN involves two key parameters: \textbf{$R$} (or often denoted as $\epsilon$) and \textbf{$MinPts$}. $R$ defines the maximum distance between two points for them to be considered neighbours. $MinPts$ specifies the minimum number of points required within the $R$-neighbourhood for a given point to be classified as a \textit{core point} - which can be thought of as the \textit{starting point} of a cluster. A point $q$ is considered \textit{directly density-reachable} from a core point $p$, if $q$ lies within the $R$-distance of $p$. A point $q$ is \textit{density-reachable} from a point $p$ if there exist a sequence of points $p_{1},\dots,p_{n}$ such that each $p_{i+1}$ is directly density-reachable from $p_{i}$, and $q$ and $p$ are at opposite ends of the chain. This means that all intermediate points in $p_{1},\dots,p_{n}$ must be core points. Finally, two points, $p$ and $q$ are \textit{density-connected} if there exists a core point $p^\prime$, from which both $p$ and $q$ are density-reachable. All points in a cluster are density-connected. It is useful to note that the classification of both core and noise points is deterministic, given a particular set of $R$ and $MinPts$ (some clustering algorithms are stochastic in nature and thus return different results if run multiple times). A smaller $R$ defines a tighter neighbourhood, potentially leading to a significant portion of legitimate cluster points being classified as noise. Conversely, a larger $R$ expands the neighbourhood radius, potentially causing distinct clusters to be counted as a single cluster. A higher $MinPts$ imposes a stricter density requirement, potentially leading to fewer, highly dense clusters. Conversely, a lower $MinPts$ relaxes the density requirement, potentially allowing noise points to form their own clusters. We trialled a range of values of $R$ and $MinPts$ and assessed the quality of the resulting clusters based on \textit{Intra-cluster Distance}, \textit{Inter-cluster Distance} and \textit{Noise Proportion}, aiming to minimise intra-cluster distance, maximise inter-cluster distances, and minimise noise proportion. Based on our experimentation, we determined that $R=0.35$ and $MinPts= 26,22,14,5$, for $m=2,4,8,16$ (or $max(5,30-2m)$).

\subsection{Clustering in DBSCAN and its interpretation in relation to common ground}

Recall that $x_{i,k}(t)$ is agent $i$'s certainty that information $k$ belongs to the common ground. Thus, at any arbitrary but fixed time $t$, the Euclidean distance $\Vert x_{i}(t) - x_j(t)\Vert$ offers a suitable measure for how similar agent~$i$ and agent~$j$'s perception of the common ground is, and more importantly, allows us to say whether they in fact share a common ground. In more detail, the model dynamics are such that the sender, agent~$i$, has an increasing probability of selecting information $k$ as $x_{i,k}(t)$ increases. Similarly, the receiver, agent~$j$, has an increasing probability of accepting the transmission as $x_{j,k}(t)$ increases. Generally speaking, if $\Vert x_{i}(t) - x_j(t)\Vert$ is small, any piece of information likely to be selected for transmission by agent $i$ will also likely be accepted~\footnote{The exception to this is if every entry of $x_i(t)$ and $x_j(t)$ is close to $-1$, but this is almost never observed in our simulations}. In other words, if we ignore the outcome in which the sender does not receive the response, then most of the grounding attempts between agents~$i$ and $j$ will be successful. Agents~$i$ and $j$ can therefore easily share information with one another, and have an established and shared common ground. In our work, we interpret DBSCAN-detected clusters as groups of agents who have successfully established a common ground. An important note is that DBSCAN detects clusters based on $x_i$ alone and does not consider connectivity. This implies that a DBSCAN-detected cluster could consist of multiple disconnected components. However, we performed additional analysis (described in the sequel) which confirms that clusters detected via DBSCAN are overwhelmingly connected in the network. In other words, the clustering corresponds to, within the network, connected groups of agents who have a shared common ground. This provides confidence in the interpretation of the results within the main paper. 

\begin{figure}[!ht]
    \centering
    \subfloat[Offline]{\centering
        \includegraphics[width=\linewidth]{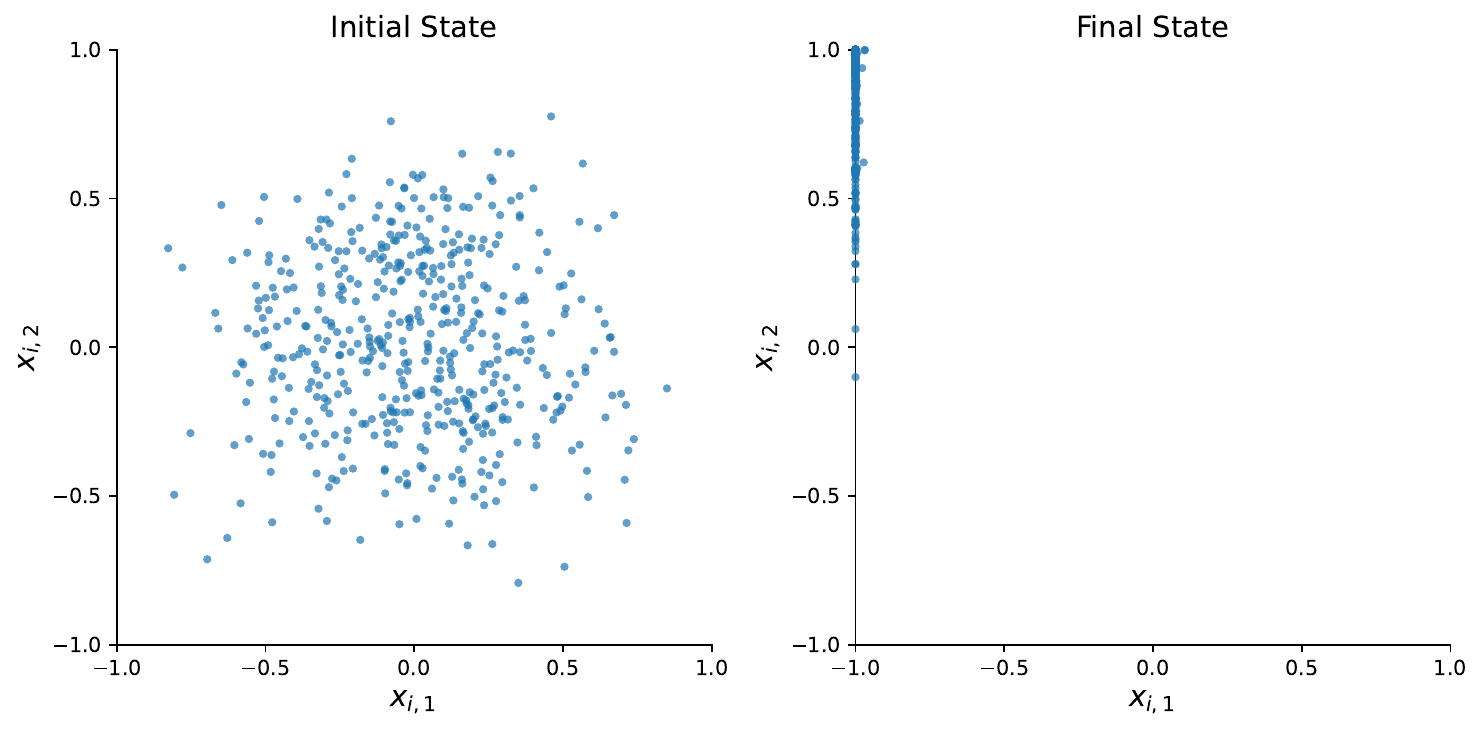}
        \label{fig:Development_offline}}
        \vfill
        \subfloat[Online]{
        \centering
        \includegraphics[width=\linewidth]{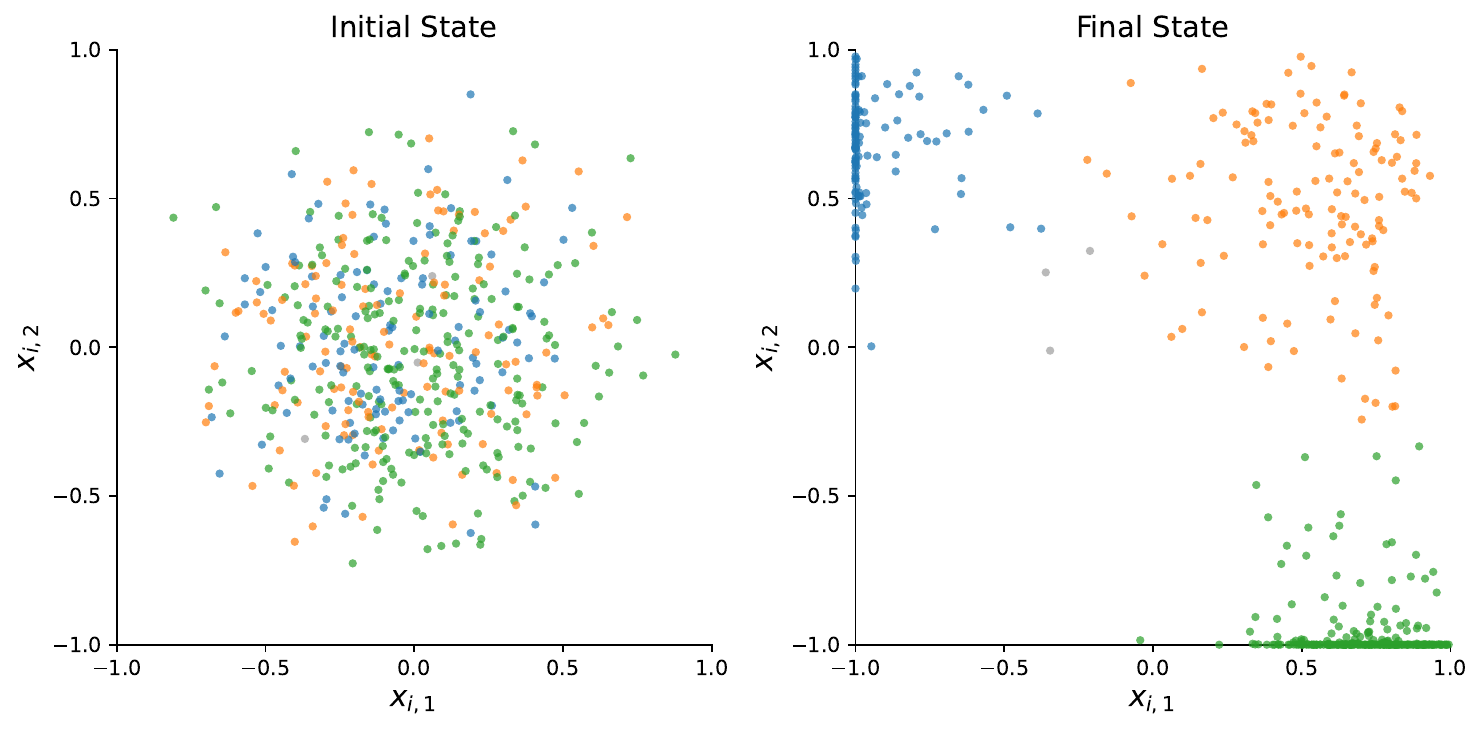}
        \label{fig:Development_online}}    
    \caption{The state vectors of agents plotted in Euclidean space, corresponding to simulations in Fig 2a (Offline) and Fig 2b (Online) of the main paper. Left panel shows $t=0$, and right panel $t=1\times 10^6$. Each node is colour-coded based on their cluster membership at $t=1\times 10^6$. For example, all nodes in cluster $k$ at $t=1\times 10^6$ have the same colour in both the initial and final state plots.}
    \label{fig:Development}
\end{figure}

Fig.~\ref{fig:Development} presents state vectors $x_i$ in $2D$ Euclidean space, for the simulation detailed in Fig.~2 of the main paper. Fig.~\ref{fig:Development} helps to visualise how DBSCAN works to group clusters, and how we can interpret agents within a cluster as sharing a common ground. The left panel of Fig.~\ref{fig:Development_offline} and \ref{fig:Development_online} show the state vectors at $t = 0$. Here, no cluster is detected, and one can see the agents are randomly distributed (in fact following the method we described in the main paper and the section above). The right panels of Fig.~\ref{fig:Development_offline} and \ref{fig:Development_online} show $t = T$, i.e. at the simulation stopping point. In the right panel of Fig.~\ref{fig:Development_offline}, there is a single cluster at $x_{i,2} \approx 1$ and $x_{i,1} \approx -1$. In other words, nearly all agents accept $x_{i,2}$ and reject $x_{i,1}$, and this is consistent with Fig.~2a in the main paper. In the right panel of Fig.~\ref{fig:Development_online}, we observe that agents form three distinct ``clusters" in the top-left, top-right and bottom-right corners (we report these in Section~III-C of the main paper). These clusters are identified by DBSCAN, and we can see how, in $2D$ space, each cluster is distinct and separated from each other.


\subsection{Noise in DBSCAN}

In DBSCAN, a \textit{noise point} is a data point whose local neighbourhood is too sparse under the chosen parameters, $R$ and $MinPts$, to be grouped into a cluster. Within the context of common ground, noise points correspond to agents who fail to share a common ground with a sufficiently large number of others. Such agents can be thought of as living in isolated information bubbles (or at most sharing a common ground with a very few number of others, below the threshold $MinPts$ to form a cluster). A population with a high proportion of noise points at termination indicates that many agents have not established common ground or that clusters have yet to form. In Fig. 5 of the main paper, we grey out regions with high proportion of noise (more than $30\%$ of the agents), as the corresponding statistics (average number of clusters, confidence intervals) offer no interpretative value. When there are high levels of noise, this means that there are many agents for whom their state vectors $x_i$ are not close together in Euclidean space. We described above how we can interpret two agents $i$ and $j$ as sharing a common ground if $x_i$ and $x_j$ are close in Euclidean space. It is therefore evident that with high levels of noise, many agents do not share a common ground with any other agent, thus we cannot discuss outcomes such as the number of clusters.


\clearpage
\section{Additional analysis for Fig.~4 and Fig.~5 of main paper}

The quantities reported in this section were collected from the set of replicated simulation runs for each triplet of $(m,\varepsilon,\gamma)$. Any values in parentheses represent the 95\% confidence interval.

The following quantities are reported.
\begin{itemize}
    \item $N_{cluster}$: Mean number of clusters identified by DBSCAN, at the end of the simulation time window.
    \item $P_{connected} = \frac{1}{K}\sum_{i=1}^K \frac{|L_i|}{|C_i|}$, where $K$ is the total number of clusters, $|L_i|$ is the number of agents in the largest connected component (within the network) of cluster $i$ and $|C_i|$ is the total number of agents in cluster $i$. This captures the proportion of agents which belong to the largest connected component of the relevant cluster as identified by DBSCAN.
    \item Inter-cluster distance = $\frac{2}{K(K-1)} \sum_{1 \le i < j \le K} \|\mu_i - \mu_j\|_2$, where $\mu_i$ is the centroid vector of cluster~$i$. An associated confidence interval is only computed if there are at least 2 simulation runs with 2 or more clusters detected. This distance describes how far apart the clusters are to each other.
    \item Intra-cluster distance = $\frac{1}{K} \sum_{k=1}^{K} \left( \frac{1}{|C_k|} \sum_{x \in C_k} \|x - \mu_k\|_2 \right)$. This distance describes how close points within a cluster are to each other. A large inter-cluster distance and a small inter-cluster distance corresponds to clear, and distinct clusters with minimal overlap in Euclidean space.
    \item Noise: The mean percentage of agents associated with noise points in DBSCAN, which are not assigned to a cluster.
\end{itemize}

\subsection{Figure~4}


To support Fig.~4 of the main paper, we report statistics of the clustering analysis for the offline scenario (Table~\ref{tab:cluster_stats_offline}) and online scenario (Table~\ref{tab:cluster_stats_online}). The data indicates that the inter-cluster distances are an order of magnitude greater than the intra-cluster distances, with low variance. In other words, the clusters are distinct and spatially separated. Moreover, $P_{connected}$ are all greater than $0.95$, except $m=16$ in the online case, where it is just below $0.9$. This indicates that the clusters identified by DBSCAN form a large connected component in the network. Overall, this indicates that that each cluster is a connected over the network and share a clear perception of the common ground, with the different clusters distinct from each other.

\begin{figure}
    \centering
   \subfloat[$m=2$]{\includegraphics[width=0.48\linewidth]{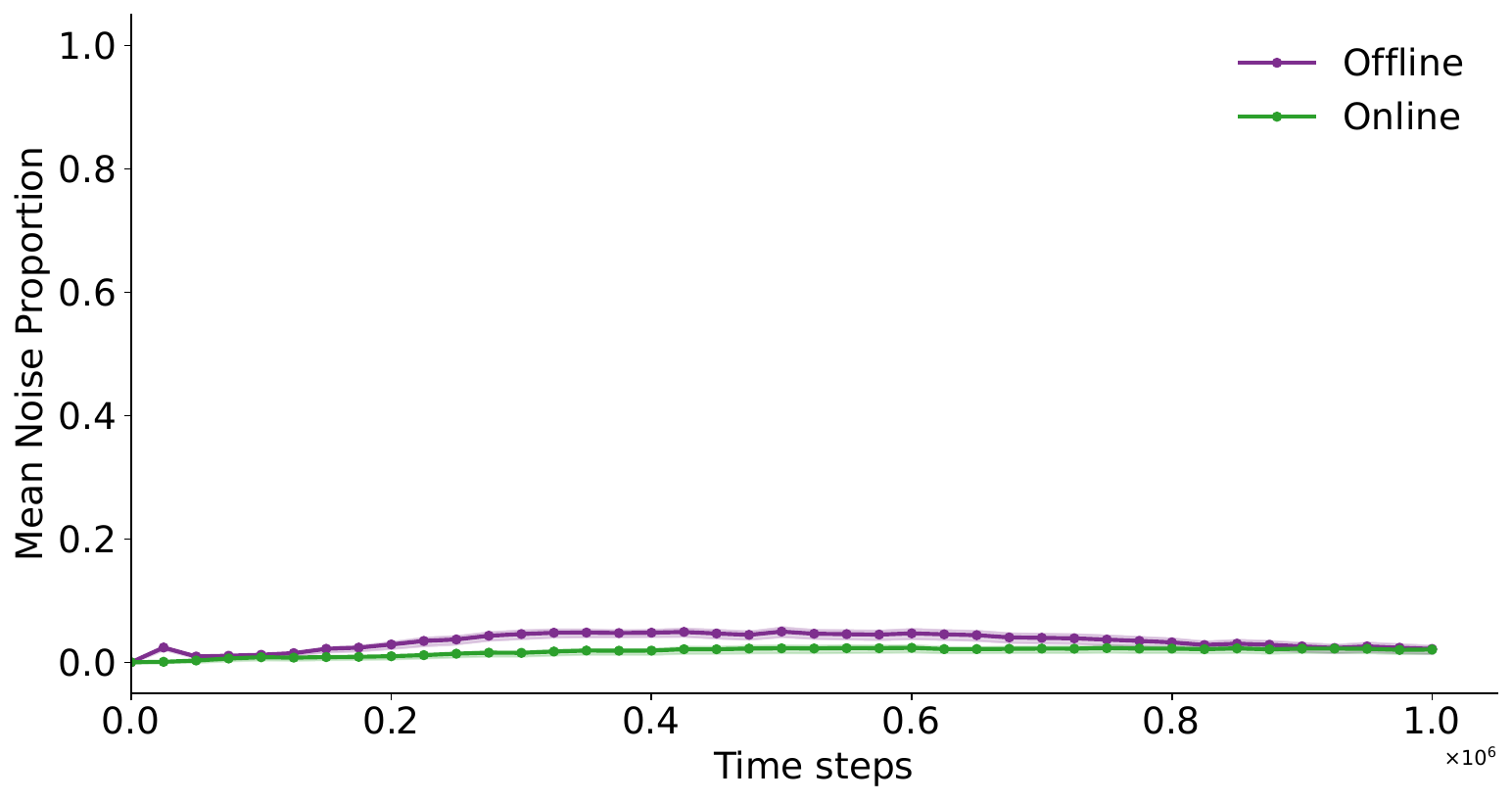}\label{fig:noise_timeseries_m2}}
    \hfill
    \subfloat[$m=4$]{\includegraphics[width=0.48\linewidth]{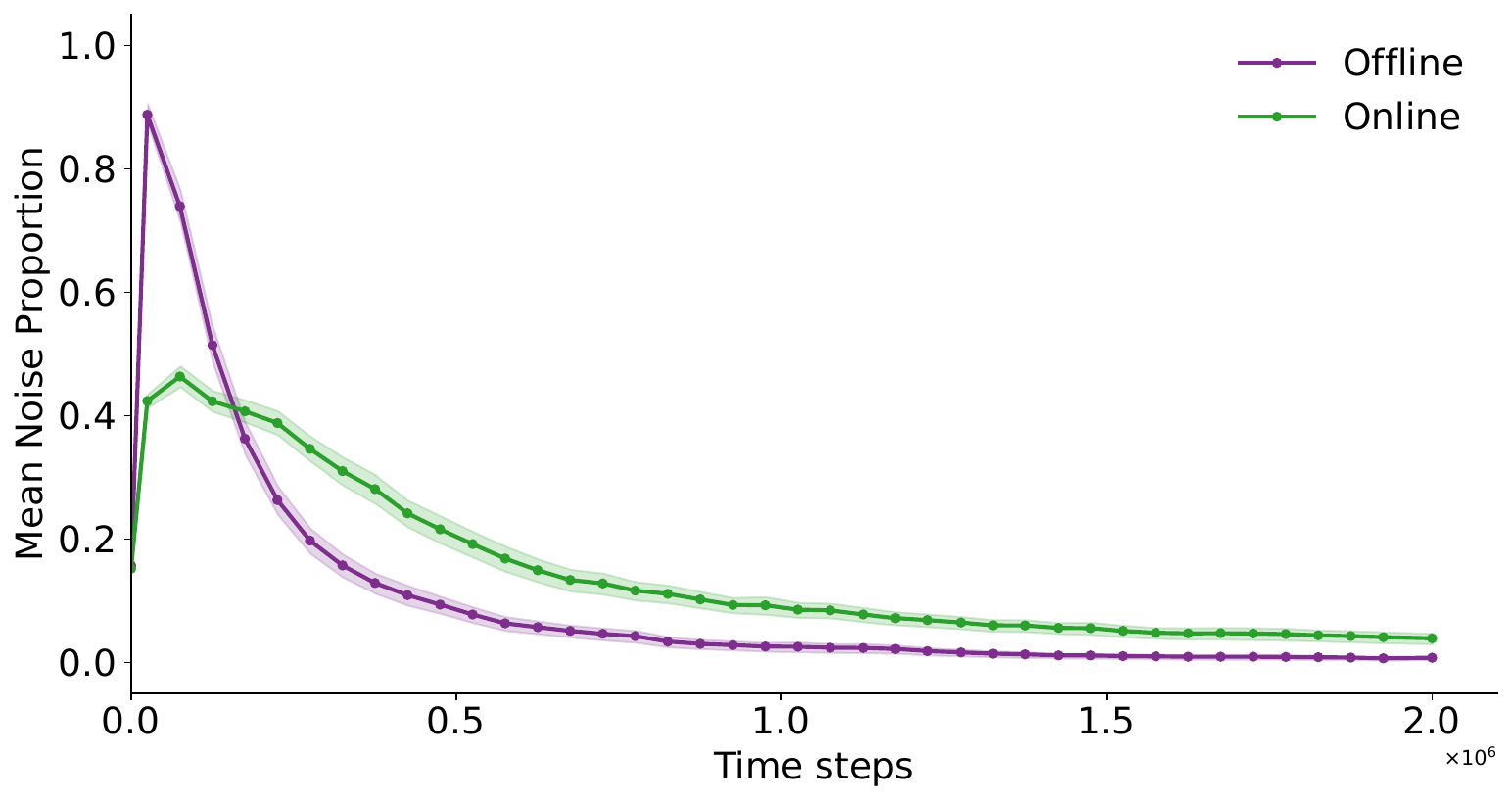}        \label{fig:cluster_timeseries_m4}}
\vfill
   \subfloat[$m=8$]{\includegraphics[width=0.48\linewidth]{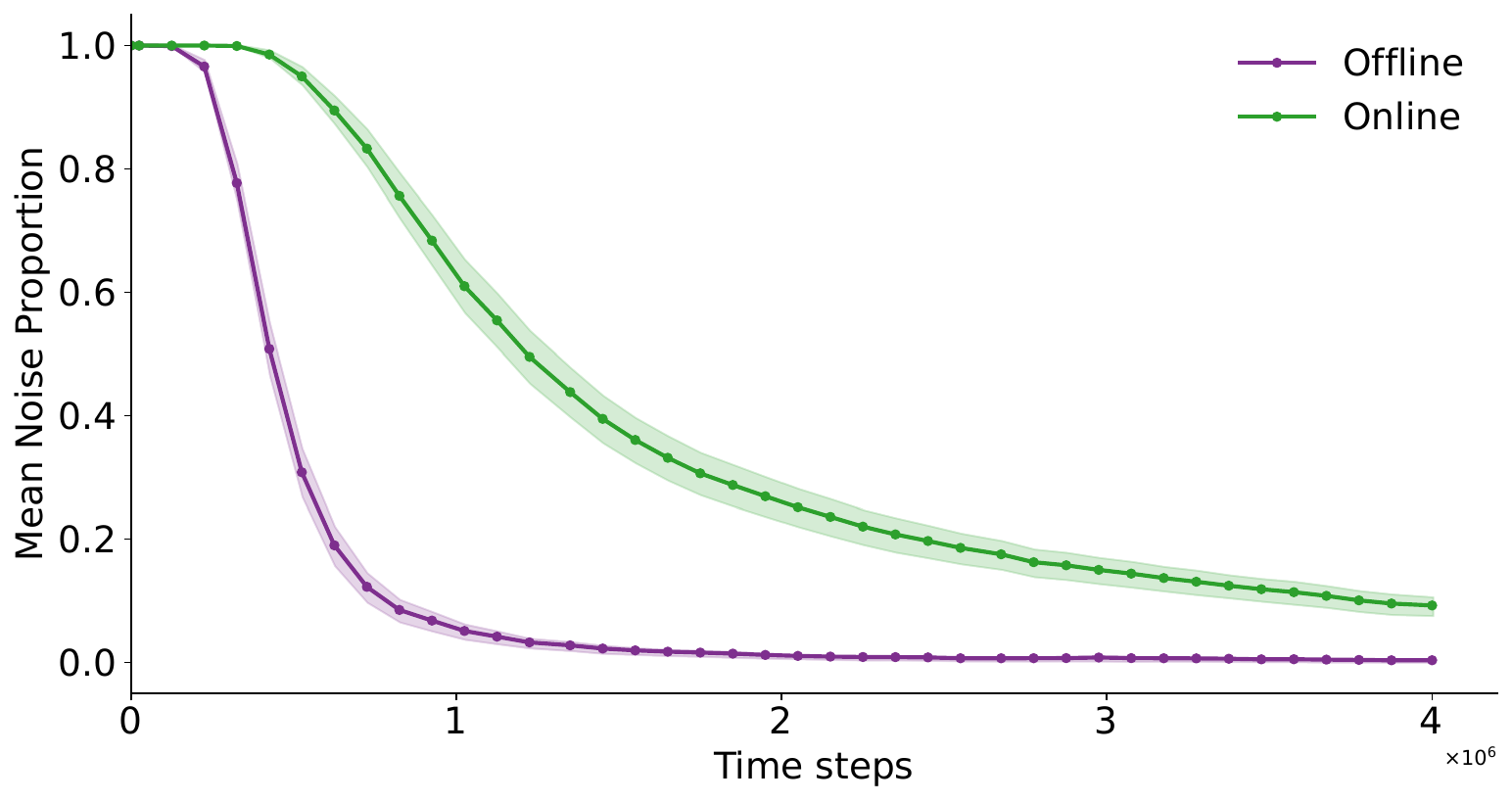}
        \label{fig:noise_timeseries_m8}}
    \hfill
    \subfloat[$m=16$]{\includegraphics[width=0.48\linewidth]{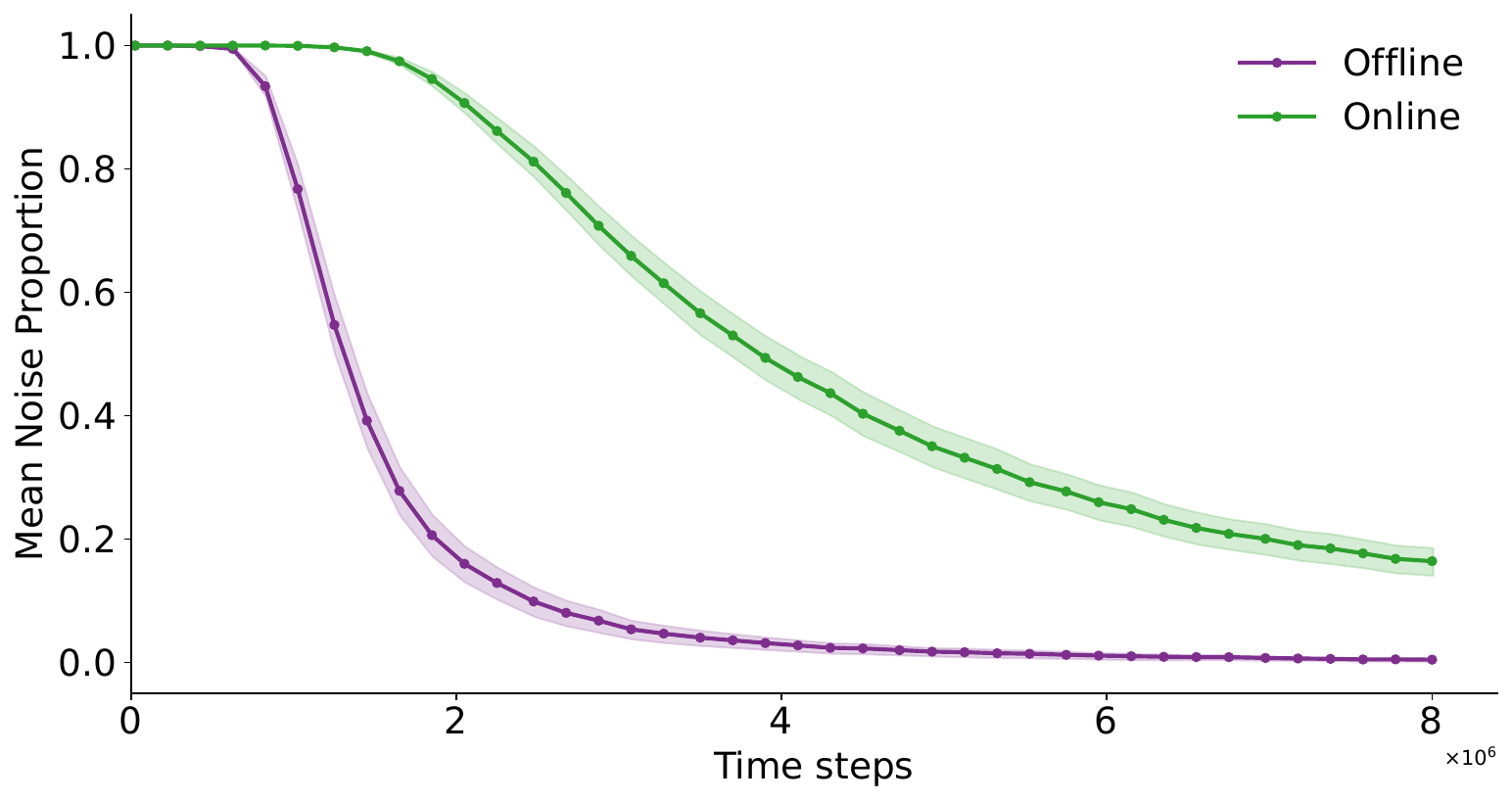}\label{fig:noise_timeseries_m16}}
    \caption{Time series of the number of noise points for offline ($\gamma = -1$ and $\varepsilon = 0.1$) and online ($\gamma = 0$ and $\varepsilon = 0.5$) settings at various values of $m$. The solid line indicates the mean of 100 Monte Carlo simulations, with the shaded region being a 95\% confidence interval. }
    \label{fig:noise_timeseries}
\end{figure}

Figure~\ref{fig:noise_timeseries} reports the fraction of nodes associated with a noise point in DBSCAN, over time. Notice that while the level of noise can initially be as high as $1$ (every point is a noise point), the noise reduces over time until at the end of the simulation, where there is less than $20\%$ of the nodes labelled as noise points. This provides additional context and support for the interpretation of the results in Section~IV-A and IV-B.

\begin{table}[htbp]
\centering
\caption{Cluster statistics for offline scenario ($\gamma=-1.0, \varepsilon=0.1$).}
\label{tab:cluster_stats_offline}
\begin{tabular}{lrrrrrr}
\toprule
\textbf{Parameters} & \textbf{reps} & \textbf{$N_{clusters}$} & $\textbf{P}_{connected}$ & \textbf{Inter-cluster} & \textbf{Intra-cluster} & \textbf{Total Var} \\
\midrule
$m=2$  & 100 & 1.2  & 0.999 & 2.36 & 0.113 & 0.0146 \\
$m=4$  & 100 & 1.3  & 0.994 & 2.03 & 0.205 & 0.0893 \\
$m=8$  & 100 & 1.06 & 0.999 & 1.83 & 0.165 & 0.0798 \\
$m=16$ & 100 & 1.42 & 0.997 & 1.89 & 0.463 & 0.249  \\
\bottomrule
\end{tabular}
\end{table}

\begin{table}[htbp]
\centering
\caption{Cluster statistics for online settings ($\gamma=0.0, \varepsilon=0.5$).}
\label{tab:cluster_stats_online}
\begin{tabular}{lrrrrrr}
\toprule
\textbf{Parameters} & \textbf{reps} & \textbf{$N_{clusters}$} & $\textbf{P}_{connected}$ & \textbf{Inter-cluster} & \textbf{Intra-cluster} & \textbf{Total Var} \\
\midrule
$m=2$  & 100 & 1.72 & 0.982 & 1.86 & 0.193 & 0.0481 \\
$m=4$  & 100 & 2.17 & 0.981 & 1.68 & 0.273 & 0.0851 \\
$m=8$  & 100 & 3.77 & 0.95  & 1.85 & 0.434 & 0.142  \\
$m=16$ & 100 & 11.2 & 0.887 & 2.28 & 0.559 & 0.104  \\
\bottomrule
\end{tabular}
\end{table}

\clearpage

\subsection{Figure~5}
Tables~\ref{tab:m2}, \ref{tab:m4}, \ref{tab:m8}, \ref{tab:m16} report additional statistical information on the DBSCAN results for the heatmaps appearing as Fig.~5 in the main paper. Of particular note is the noise column, which provides context for why, in the $m=8$ and $m=16$ plots in Fig~5, we greyed out certain cells that had high levels of noise.

\begin{table}[htbp]
  \centering
  \caption{Cluster summary for $m = 2$}
  \label{tab:m2}
  \footnotesize
  \setlength{\tabcolsep}{4pt}
  \begin{tabular}{cc ccccr}
    \toprule
    $\varepsilon$ & $\gamma$ & $N_{\text{cluster}}$ & $P_{\text{connected}}$ & Inter-cluster distance & Intra-cluster distance & Noise (\%) \\
    \midrule
    0.1 & $-1$ & 1.2 (1.12, 1.28) & 0.999 (0.997, 1) & 2.36 (2.3, 2.43) & 0.113 (0.103, 0.123) & 1.68 \\
    0.1 & $-0.8$ & 1.6 (1.45, 1.75) & 0.987 (0.983, 0.992) & 2.17 (2.07, 2.27) & 0.136 (0.123, 0.149) & 3.63 \\
    0.1 & $-0.6$ & 2 (1.82, 2.18) & 0.97 (0.962, 0.977) & 2.02 (1.94, 2.09) & 0.15 (0.135, 0.165) & 3.75 \\
    0.1 & $-0.4$ & 2.09 (1.92, 2.26) & 0.968 (0.96, 0.975) & 2.05 (1.98, 2.12) & 0.161 (0.138, 0.184) & 3.91 \\
    0.1 & $-0.2$ & 2.15 (1.96, 2.34) & 0.97 (0.963, 0.977) & 1.94 (1.89, 1.99) & 0.135 (0.12, 0.15) & 3.21 \\
    0.1 & $0$ & 2.41 (2.24, 2.58) & 0.964 (0.957, 0.971) & 1.99 (1.95, 2.03) & 0.176 (0.152, 0.2) & 2.67 \\
    0.1 & $+0.2$ & 2.21 (2.03, 2.39) & 0.972 (0.965, 0.979) & 2 (1.95, 2.06) & 0.152 (0.127, 0.177) & 3.14 \\
    0.1 & $+0.4$ & 2.34 (2.17, 2.51) & 0.965 (0.959, 0.972) & 1.97 (1.92, 2.03) & 0.148 (0.126, 0.17) & 2.84 \\
    0.1 & $+0.6$ & 2.42 (2.26, 2.58) & 0.967 (0.961, 0.973) & 1.97 (1.92, 2.02) & 0.152 (0.128, 0.176) & 3.04 \\
    0.1 & $+0.8$ & 2.12 (1.93, 2.31) & 0.973 (0.966, 0.98) & 2.02 (1.97, 2.07) & 0.119 (0.101, 0.137) & 3.33 \\
    0.1 & $+1$ & 2.38 (2.21, 2.55) & 0.968 (0.962, 0.975) & 2.05 (2, 2.09) & 0.16 (0.133, 0.188) & 2.45 \\
    \addlinespace[3pt]
    0.2 & $-1$ & 1 (1, 1) & 1 (1, 1) & --- & 0.131 (0.121, 0.142) & 0.01 \\
    0.2 & $-0.8$ & 1 (1, 1) & 1 (1, 1) & --- & 0.114 (0.113, 0.115) & 0.05 \\
    0.2 & $-0.6$ & 1.25 (1.16, 1.34) & 0.999 (0.998, 1) & 2.32 (2.3, 2.33) & 0.131 (0.12, 0.141) & 2.31 \\
    0.2 & $-0.4$ & 1.69 (1.52, 1.86) & 0.983 (0.978, 0.989) & 2.03 (1.93, 2.12) & 0.143 (0.129, 0.157) & 3.47 \\
    0.2 & $-0.2$ & 1.78 (1.6, 1.96) & 0.978 (0.973, 0.984) & 1.92 (1.84, 2) & 0.127 (0.116, 0.138) & 3.53 \\
    0.2 & $0$ & 2.24 (2.06, 2.42) & 0.971 (0.965, 0.977) & 1.93 (1.88, 1.97) & 0.156 (0.138, 0.175) & 3.22 \\
    0.2 & $+0.2$ & 2.07 (1.88, 2.26) & 0.969 (0.961, 0.976) & 1.93 (1.88, 1.98) & 0.137 (0.115, 0.159) & 2.78 \\
    0.2 & $+0.4$ & 2.26 (2.08, 2.44) & 0.968 (0.962, 0.975) & 1.97 (1.91, 2.02) & 0.145 (0.123, 0.166) & 3.20 \\
    0.2 & $+0.6$ & 2.44 (2.28, 2.6) & 0.971 (0.964, 0.978) & 1.96 (1.93, 2) & 0.168 (0.141, 0.194) & 2.02 \\
    0.2 & $+0.8$ & 2.14 (1.96, 2.32) & 0.972 (0.965, 0.98) & 2.02 (1.97, 2.07) & 0.165 (0.13, 0.2) & 2.92 \\
    0.2 & $+1$ & 2.38 (2.23, 2.53) & 0.971 (0.965, 0.977) & 1.99 (1.94, 2.04) & 0.183 (0.146, 0.219) & 2.30 \\
    \addlinespace[3pt]
    0.3 & $-1$ & 1 (1, 1) & 1 (1, 1) & --- & 0.058 (0.0569, 0.059) & 0.01 \\
    0.3 & $-0.8$ & 1 (1, 1) & 1 (1, 1) & --- & 0.111 (0.105, 0.117) & 0.06 \\
    0.3 & $-0.6$ & 1.02 (0.992, 1.05) & 1 (1, 1) & 1.96 (1.76, 2.16) & 0.13 (0.124, 0.135) & 0.07 \\
    0.3 & $-0.4$ & 1.4 (1.27, 1.53) & 0.994 (0.991, 0.997) & 2.03 (1.93, 2.14) & 0.144 (0.133, 0.155) & 2.56 \\
    0.3 & $-0.2$ & 1.83 (1.65, 2.01) & 0.978 (0.972, 0.984) & 1.92 (1.84, 1.99) & 0.151 (0.135, 0.167) & 3.54 \\
    0.3 & $0$ & 1.88 (1.69, 2.07) & 0.981 (0.976, 0.987) & 1.86 (1.81, 1.92) & 0.157 (0.135, 0.178) & 2.74 \\
    0.3 & $+0.2$ & 2.09 (1.92, 2.26) & 0.975 (0.968, 0.981) & 1.95 (1.89, 2) & 0.188 (0.155, 0.22) & 2.46 \\
    0.3 & $+0.4$ & 2.21 (2.05, 2.37) & 0.972 (0.967, 0.978) & 1.93 (1.88, 1.99) & 0.216 (0.174, 0.258) & 2.48 \\
    0.3 & $+0.6$ & 2.38 (2.22, 2.54) & 0.97 (0.964, 0.976) & 1.97 (1.93, 2.02) & 0.208 (0.17, 0.245) & 1.87 \\
    0.3 & $+0.8$ & 2.17 (2.01, 2.33) & 0.975 (0.97, 0.98) & 1.96 (1.9, 2.02) & 0.179 (0.141, 0.218) & 2.62 \\
    0.3 & $+1$ & 2.36 (2.19, 2.53) & 0.977 (0.973, 0.982) & 2.06 (2.01, 2.1) & 0.205 (0.167, 0.243) & 1.99 \\
    \addlinespace[3pt]
    0.4 & $-1$ & 1 (1, 1) & 1 (1, 1) & --- & 0.0553 (0.0542, 0.0563) & 0.01 \\
    0.4 & $-0.8$ & 1 (1, 1) & 1 (1, 1) & --- & 0.0964 (0.095, 0.0979) & 0.02 \\
    0.4 & $-0.6$ & 1 (1, 1) & 1 (1, 1) & --- & 0.162 (0.161, 0.164) & 0.12 \\
    0.4 & $-0.4$ & 1.07 (1.02, 1.12) & 1 (1, 1) & 2.07 (2.03, 2.11) & 0.139 (0.132, 0.146) & 0.91 \\
    0.4 & $-0.2$ & 1.81 (1.63, 1.99) & 0.976 (0.968, 0.983) & 1.8 (1.74, 1.87) & 0.173 (0.153, 0.194) & 2.25 \\
    0.4 & $0$ & 1.92 (1.74, 2.1) & 0.974 (0.968, 0.981) & 1.84 (1.78, 1.9) & 0.166 (0.141, 0.191) & 2.54 \\
    0.4 & $+0.2$ & 1.9 (1.74, 2.06) & 0.978 (0.972, 0.984) & 1.87 (1.8, 1.94) & 0.203 (0.164, 0.241) & 2.92 \\
    0.4 & $+0.4$ & 2.01 (1.84, 2.18) & 0.979 (0.974, 0.984) & 1.89 (1.83, 1.95) & 0.212 (0.17, 0.254) & 2.26 \\
    0.4 & $+0.6$ & 2.19 (2.02, 2.36) & 0.978 (0.973, 0.982) & 1.97 (1.93, 2.02) & 0.215 (0.174, 0.256) & 2.20 \\
    0.4 & $+0.8$ & 2.36 (2.2, 2.52) & 0.975 (0.97, 0.98) & 2.04 (2.01, 2.08) & 0.25 (0.204, 0.295) & 1.87 \\
    0.4 & $+1$ & 2.39 (2.23, 2.55) & 0.979 (0.975, 0.983) & 2.06 (2.01, 2.1) & 0.23 (0.19, 0.27) & 1.61 \\
    \addlinespace[3pt]
    0.5 & $-1$ & 1 (1, 1) & 1 (1, 1) & --- & 0.0503 (0.0493, 0.0513) & 0.00 \\
    0.5 & $-0.8$ & 1 (1, 1) & 1 (1, 1) & --- & 0.097 (0.0958, 0.0982) & 0.03 \\
    0.5 & $-0.6$ & 1 (1, 1) & 1 (1, 1) & --- & 0.17 (0.169, 0.172) & 0.17 \\
    0.5 & $-0.4$ & 1.01 (0.99, 1.03) & 1 (1, 1) & 1.88 (---) & 0.157 (0.154, 0.16) & 0.28 \\
    0.5 & $-0.2$ & 1.48 (1.33, 1.63) & 0.983 (0.976, 0.989) & 1.79 (1.71, 1.87) & 0.166 (0.15, 0.183) & 2.11 \\
    0.5 & $0$ & 1.72 (1.56, 1.88) & 0.982 (0.976, 0.987) & 1.86 (1.79, 1.94) & 0.193 (0.16, 0.225) & 2.39 \\
    0.5 & $+0.2$ & 1.81 (1.64, 1.98) & 0.987 (0.983, 0.991) & 1.84 (1.79, 1.89) & 0.254 (0.207, 0.302) & 2.31 \\
    0.5 & $+0.4$ & 2.26 (2.1, 2.42) & 0.977 (0.973, 0.982) & 1.92 (1.88, 1.96) & 0.268 (0.222, 0.313) & 1.58 \\
    0.5 & $+0.6$ & 2.27 (2.1, 2.44) & 0.979 (0.975, 0.984) & 1.93 (1.89, 1.97) & 0.229 (0.188, 0.269) & 1.77 \\
    0.5 & $+0.8$ & 2.41 (2.28, 2.54) & 0.973 (0.967, 0.979) & 1.95 (1.91, 1.98) & 0.207 (0.17, 0.243) & 1.81 \\
    0.5 & $+1$ & 2.45 (2.32, 2.58) & 0.971 (0.965, 0.977) & 2.01 (1.97, 2.05) & 0.187 (0.158, 0.216) & 1.83 \\
    \bottomrule
  \end{tabular}
\end{table}

\clearpage

\begin{table}[htbp]
  \centering
  \caption{Cluster summary for $m = 4$}
  \label{tab:m4}
  \footnotesize
  \setlength{\tabcolsep}{4pt}
  \begin{tabular}{cc ccccr}
    \toprule
    $\varepsilon$ & $\gamma$ & $N_{\text{cluster}}$ & $P_{\text{connected}}$ & Inter-cluster distance & Intra-cluster distance & Noise (\%) \\
    \midrule
    0.1 & $-1$ & 1.3 (1.18, 1.42) & 0.994 (0.991, 0.998) & 2.03 (1.93, 2.14) & 0.205 (0.15, 0.26) & 1.00 \\
    0.1 & $-0.8$ & 1.53 (1.4, 1.66) & 0.992 (0.989, 0.995) & 2.03 (1.95, 2.11) & 0.284 (0.229, 0.339) & 2.51 \\
    0.1 & $-0.6$ & 1.74 (1.54, 1.94) & 0.984 (0.979, 0.99) & 1.98 (1.9, 2.06) & 0.234 (0.186, 0.283) & 4.30 \\
    0.1 & $-0.4$ & 2.06 (1.83, 2.29) & 0.981 (0.976, 0.987) & 1.99 (1.93, 2.06) & 0.252 (0.206, 0.299) & 5.20 \\
    0.1 & $-0.2$ & 2.48 (2.19, 2.77) & 0.964 (0.954, 0.974) & 2.07 (2.02, 2.12) & 0.277 (0.231, 0.322) & 8.13 \\
    0.1 & $0$ & 2.45 (2.18, 2.72) & 0.969 (0.961, 0.977) & 2.05 (2, 2.11) & 0.25 (0.205, 0.295) & 8.27 \\
    0.1 & $+0.2$ & 2.74 (2.42, 3.06) & 0.96 (0.949, 0.97) & 2.08 (2.03, 2.13) & 0.323 (0.272, 0.373) & 8.66 \\
    0.1 & $+0.4$ & 2.86 (2.54, 3.18) & 0.956 (0.945, 0.967) & 2.07 (2.02, 2.13) & 0.265 (0.222, 0.308) & 9.01 \\
    0.1 & $+0.6$ & 3.46 (3.1, 3.82) & 0.937 (0.924, 0.951) & 2.14 (2.09, 2.19) & 0.239 (0.207, 0.27) & 10.70 \\
    0.1 & $+0.8$ & 2.84 (2.56, 3.12) & 0.958 (0.947, 0.969) & 2.07 (2.02, 2.13) & 0.283 (0.241, 0.325) & 10.90 \\
    0.1 & $+1$ & 3.45 (3.11, 3.79) & 0.937 (0.924, 0.95) & 2.15 (2.1, 2.2) & 0.235 (0.211, 0.26) & 11.70 \\
    \addlinespace[3pt]
    0.2 & $-1$ & 1 (1, 1) & 1 (1, 1) & --- & 0.0774 (0.0764, 0.0784) & 0.00 \\
    0.2 & $-0.8$ & 1 (1, 1) & 1 (1, 1) & --- & 0.0962 (0.0941, 0.0984) & 0.00 \\
    0.2 & $-0.6$ & 1.12 (1.05, 1.19) & 0.998 (0.997, 1) & 2.03 (1.94, 2.11) & 0.157 (0.112, 0.203) & 0.52 \\
    0.2 & $-0.4$ & 1.59 (1.44, 1.74) & 0.99 (0.986, 0.993) & 1.9 (1.83, 1.98) & 0.265 (0.21, 0.32) & 2.54 \\
    0.2 & $-0.2$ & 1.69 (1.49, 1.89) & 0.986 (0.981, 0.991) & 1.88 (1.81, 1.96) & 0.223 (0.175, 0.27) & 3.69 \\
    0.2 & $0$ & 2.21 (1.91, 2.51) & 0.978 (0.971, 0.986) & 1.99 (1.93, 2.06) & 0.312 (0.259, 0.364) & 5.82 \\
    0.2 & $+0.2$ & 2.67 (2.37, 2.97) & 0.965 (0.955, 0.974) & 1.99 (1.94, 2.03) & 0.284 (0.242, 0.327) & 8.07 \\
    0.2 & $+0.4$ & 3.08 (2.77, 3.39) & 0.956 (0.946, 0.967) & 2.07 (2.03, 2.12) & 0.272 (0.236, 0.308) & 8.52 \\
    0.2 & $+0.6$ & 3.27 (2.92, 3.62) & 0.949 (0.937, 0.962) & 2.03 (1.97, 2.08) & 0.265 (0.234, 0.296) & 9.48 \\
    0.2 & $+0.8$ & 3.6 (3.3, 3.9) & 0.943 (0.932, 0.955) & 2.07 (2.02, 2.11) & 0.233 (0.217, 0.249) & 12.00 \\
    0.2 & $+1$ & 4.09 (3.81, 4.37) & 0.93 (0.918, 0.941) & 2.13 (2.09, 2.17) & 0.241 (0.225, 0.256) & 12.00 \\
    \addlinespace[3pt]
    0.3 & $-1$ & 1 (1, 1) & 1 (1, 1) & --- & 0.0733 (0.0723, 0.0743) & 0.00 \\
    0.3 & $-0.8$ & 1 (1, 1) & 1 (1, 1) & --- & 0.113 (0.113, 0.114) & 0.00 \\
    0.3 & $-0.6$ & 1 (1, 1) & 1 (1, 1) & --- & 0.14 (0.134, 0.147) & 0.01 \\
    0.3 & $-0.4$ & 1.19 (1.09, 1.29) & 0.997 (0.994, 0.999) & 1.95 (1.87, 2.04) & 0.204 (0.15, 0.259) & 0.94 \\
    0.3 & $-0.2$ & 1.72 (1.52, 1.92) & 0.988 (0.983, 0.992) & 1.9 (1.83, 1.98) & 0.257 (0.209, 0.305) & 4.03 \\
    0.3 & $0$ & 2.29 (2, 2.58) & 0.974 (0.966, 0.982) & 1.91 (1.87, 1.96) & 0.287 (0.238, 0.335) & 4.84 \\
    0.3 & $+0.2$ & 2.94 (2.65, 3.23) & 0.958 (0.948, 0.969) & 1.92 (1.87, 1.96) & 0.283 (0.245, 0.321) & 6.69 \\
    0.3 & $+0.4$ & 3.1 (2.82, 3.38) & 0.956 (0.946, 0.966) & 1.92 (1.87, 1.96) & 0.242 (0.214, 0.269) & 7.71 \\
    0.3 & $+0.6$ & 3.7 (3.43, 3.97) & 0.944 (0.934, 0.954) & 2.01 (1.97, 2.05) & 0.232 (0.22, 0.245) & 9.87 \\
    0.3 & $+0.8$ & 4.18 (3.88, 4.48) & 0.918 (0.907, 0.93) & 2.16 (2.11, 2.2) & 0.248 (0.227, 0.268) & 13.60 \\
    0.3 & $+1$ & 4.23 (3.99, 4.47) & 0.923 (0.911, 0.935) & 2.19 (2.15, 2.23) & 0.236 (0.227, 0.245) & 14.20 \\
    \addlinespace[3pt]
    0.4 & $-1$ & 1 (1, 1) & 1 (1, 1) & --- & 0.0691 (0.0682, 0.07) & 0.00 \\
    0.4 & $-0.8$ & 1 (1, 1) & 1 (1, 1) & --- & 0.117 (0.117, 0.118) & 0.00 \\
    0.4 & $-0.6$ & 1 (1, 1) & 1 (1, 1) & --- & 0.186 (0.185, 0.187) & 0.01 \\
    0.4 & $-0.4$ & 1.03 (0.996, 1.06) & 0.999 (0.997, 1) & 1.95 (1.9, 2.01) & 0.13 (0.106, 0.153) & 0.15 \\
    0.4 & $-0.2$ & 1.69 (1.49, 1.89) & 0.989 (0.984, 0.993) & 1.8 (1.75, 1.85) & 0.276 (0.227, 0.325) & 3.01 \\
    0.4 & $0$ & 2.22 (1.92, 2.52) & 0.976 (0.968, 0.984) & 1.83 (1.79, 1.88) & 0.305 (0.255, 0.355) & 4.59 \\
    0.4 & $+0.2$ & 3.04 (2.74, 3.34) & 0.967 (0.96, 0.974) & 1.86 (1.82, 1.9) & 0.294 (0.253, 0.335) & 7.29 \\
    0.4 & $+0.4$ & 3.5 (3.27, 3.73) & 0.956 (0.95, 0.963) & 1.94 (1.9, 1.98) & 0.228 (0.214, 0.243) & 8.81 \\
    0.4 & $+0.6$ & 4.19 (3.87, 4.51) & 0.926 (0.913, 0.938) & 2.15 (2.1, 2.19) & 0.26 (0.237, 0.283) & 14.00 \\
    0.4 & $+0.8$ & 4.49 (4.18, 4.8) & 0.917 (0.904, 0.93) & 2.21 (2.17, 2.25) & 0.241 (0.227, 0.255) & 12.90 \\
    0.4 & $+1$ & 4.87 (4.55, 5.19) & 0.892 (0.878, 0.907) & 2.3 (2.26, 2.33) & 0.246 (0.239, 0.253) & 16.90 \\
    \addlinespace[3pt]
    0.5 & $-1$ & 1 (1, 1) & 1 (1, 1) & --- & 0.0657 (0.0649, 0.0665) & 0.00 \\
    0.5 & $-0.8$ & 1 (1, 1) & 1 (1, 1) & --- & 0.12 (0.119, 0.121) & 0.00 \\
    0.5 & $-0.6$ & 1 (1, 1) & 1 (1, 1) & --- & 0.199 (0.198, 0.2) & 0.02 \\
    0.5 & $-0.4$ & 1.01 (0.99, 1.03) & 1 (1, 1) & 1.17 (---) & 0.15 (0.142, 0.158) & 0.02 \\
    0.5 & $-0.2$ & 1.2 (1.1, 1.3) & 0.997 (0.994, 0.999) & 1.68 (1.58, 1.78) & 0.206 (0.161, 0.251) & 1.00 \\
    0.5 & $0$ & 2.17 (1.96, 2.38) & 0.981 (0.976, 0.986) & 1.68 (1.64, 1.72) & 0.273 (0.233, 0.313) & 3.66 \\
    0.5 & $+0.2$ & 3.27 (3.02, 3.52) & 0.961 (0.955, 0.966) & 1.84 (1.8, 1.88) & 0.247 (0.224, 0.269) & 7.50 \\
    0.5 & $+0.4$ & 3.97 (3.65, 4.29) & 0.936 (0.925, 0.948) & 2.02 (1.98, 2.06) & 0.266 (0.241, 0.291) & 11.60 \\
    0.5 & $+0.6$ & 4.48 (4.13, 4.83) & 0.922 (0.91, 0.933) & 2.13 (2.1, 2.16) & 0.232 (0.223, 0.241) & 12.30 \\
    0.5 & $+0.8$ & 4.75 (4.45, 5.05) & 0.904 (0.892, 0.916) & 2.18 (2.15, 2.22) & 0.24 (0.23, 0.249) & 13.00 \\
    0.5 & $+1$ & 5.07 (4.8, 5.34) & 0.89 (0.879, 0.901) & 2.31 (2.28, 2.33) & 0.246 (0.24, 0.253) & 13.80 \\
    \bottomrule
  \end{tabular}
\end{table}

\clearpage

\begin{table}[htbp]
  \centering
  \caption{Cluster summary for $m = 8$}
  \label{tab:m8}
  \footnotesize
  \setlength{\tabcolsep}{4pt}
  \begin{tabular}{cc ccccr}
    \toprule
    $\varepsilon$ & $\gamma$ & $N_{\text{cluster}}$ & $P_{\text{connected}}$ & Inter-cluster distance & Intra-cluster distance & Noise (\%) \\
    \midrule
    0.1 & $-1$ & 1.06 (0.999, 1.12) & 0.999 (0.998, 1) & 1.83 (1.67, 2) & 0.165 (0.116, 0.213) & 0.14 \\
    0.1 & $-0.8$ & 1.4 (1.25, 1.55) & 0.991 (0.988, 0.995) & 1.86 (1.79, 1.94) & 0.333 (0.265, 0.401) & 1.16 \\
    0.1 & $-0.6$ & 2.11 (1.8, 2.42) & 0.977 (0.97, 0.985) & 1.92 (1.85, 1.99) & 0.46 (0.395, 0.525) & 3.48 \\
    0.1 & $-0.4$ & 2.46 (2.15, 2.77) & 0.976 (0.969, 0.984) & 1.92 (1.87, 1.97) & 0.469 (0.409, 0.529) & 4.70 \\
    0.1 & $-0.2$ & 3.15 (2.83, 3.47) & 0.962 (0.953, 0.972) & 1.95 (1.91, 2) & 0.443 (0.388, 0.498) & 5.65 \\
    0.1 & $0$ & 4.41 (4, 4.82) & 0.933 (0.92, 0.947) & 2.11 (2.06, 2.15) & 0.378 (0.335, 0.422) & 11.00 \\
    0.1 & $+0.2$ & 4.83 (4.35, 5.31) & 0.922 (0.906, 0.937) & 2.12 (2.07, 2.17) & 0.332 (0.297, 0.368) & 12.70 \\
    0.1 & $+0.4$ & 5.01 (4.64, 5.38) & 0.91 (0.894, 0.926) & 2.2 (2.16, 2.25) & 0.333 (0.304, 0.363) & 16.70 \\
    0.1 & $+0.6$ & 5.4 (5, 5.8) & 0.906 (0.893, 0.919) & 2.25 (2.21, 2.29) & 0.353 (0.321, 0.385) & 16.60 \\
    0.1 & $+0.8$ & 5.63 (5.16, 6.1) & 0.88 (0.859, 0.9) & 2.34 (2.29, 2.39) & 0.337 (0.304, 0.37) & 23.70 \\
    0.1 & $+1$ & 5.82 (5.4, 6.24) & 0.868 (0.848, 0.887) & 2.39 (2.34, 2.44) & 0.325 (0.305, 0.345) & 27.30 \\
    \addlinespace[3pt]
    0.2 & $-1$ & 1 (1, 1) & 1 (1, 1) & --- & 0.104 (0.103, 0.104) & 0.00 \\
    0.2 & $-0.8$ & 1 (1, 1) & 1 (1, 1) & --- & 0.135 (0.132, 0.138) & 0.00 \\
    0.2 & $-0.6$ & 1.06 (1.01, 1.11) & 0.999 (0.999, 1) & 1.64 (1.61, 1.67) & 0.102 (0.0806, 0.123) & 0.20 \\
    0.2 & $-0.4$ & 1.7 (1.45, 1.95) & 0.988 (0.981, 0.994) & 1.81 (1.75, 1.87) & 0.347 (0.286, 0.409) & 1.76 \\
    0.2 & $-0.2$ & 2.4 (2.09, 2.71) & 0.976 (0.967, 0.984) & 1.87 (1.83, 1.91) & 0.455 (0.394, 0.517) & 3.58 \\
    0.2 & $0$ & 3.9 (3.52, 4.28) & 0.944 (0.933, 0.956) & 2.03 (1.98, 2.08) & 0.418 (0.367, 0.469) & 8.24 \\
    0.2 & $+0.2$ & 4.74 (4.33, 5.15) & 0.918 (0.904, 0.933) & 2.13 (2.09, 2.17) & 0.382 (0.339, 0.425) & 14.60 \\
    0.2 & $+0.4$ & 5.35 (4.92, 5.78) & 0.905 (0.889, 0.92) & 2.23 (2.19, 2.28) & 0.317 (0.289, 0.346) & 17.60 \\
    0.2 & $+0.6$ & 5.69 (5.29, 6.09) & 0.879 (0.862, 0.897) & 2.28 (2.24, 2.33) & 0.328 (0.301, 0.355) & 26.30 \\
    0.2 & $+0.8$ & 6.59 (6.22, 6.96) & 0.828 (0.807, 0.85) & 2.41 (2.37, 2.46) & 0.309 (0.294, 0.323) & 34.30 \\
    0.2 & $+1$ & 6.54 (6.19, 6.89) & 0.805 (0.784, 0.825) & 2.49 (2.44, 2.54) & 0.314 (0.301, 0.327) & 39.70 \\
    \addlinespace[3pt]
    0.3 & $-1$ & 1 (1, 1) & 1 (1, 1) & --- & 0.0989 (0.0981, 0.0997) & 0.00 \\
    0.3 & $-0.8$ & 1 (1, 1) & 1 (1, 1) & --- & 0.148 (0.148, 0.149) & 0.00 \\
    0.3 & $-0.6$ & 1 (1, 1) & 1 (1, 1) & --- & 0.214 (0.208, 0.22) & 0.00 \\
    0.3 & $-0.4$ & 1.17 (1.05, 1.29) & 0.998 (0.996, 0.999) & 1.69 (1.6, 1.79) & 0.195 (0.151, 0.239) & 0.37 \\
    0.3 & $-0.2$ & 2.12 (1.81, 2.43) & 0.979 (0.97, 0.989) & 1.81 (1.75, 1.87) & 0.445 (0.386, 0.504) & 3.64 \\
    0.3 & $0$ & 3.73 (3.31, 4.15) & 0.952 (0.942, 0.962) & 1.96 (1.92, 2.01) & 0.406 (0.357, 0.456) & 8.69 \\
    0.3 & $+0.2$ & 5.3 (4.86, 5.74) & 0.9 (0.884, 0.916) & 2.13 (2.09, 2.18) & 0.308 (0.284, 0.332) & 18.00 \\
    0.3 & $+0.4$ & 6.09 (5.7, 6.48) & 0.872 (0.854, 0.891) & 2.24 (2.2, 2.28) & 0.324 (0.303, 0.345) & 25.10 \\
    0.3 & $+0.6$ & 6.65 (6.33, 6.97) & 0.827 (0.808, 0.846) & 2.39 (2.35, 2.43) & 0.307 (0.295, 0.319) & 34.70 \\
    0.3 & $+0.8$ & 6.62 (6.31, 6.93) & 0.816 (0.799, 0.834) & 2.43 (2.39, 2.47) & 0.297 (0.292, 0.302) & 38.30 \\
    0.3 & $+1$ & 6.36 (5.98, 6.74) & 0.775 (0.753, 0.797) & 2.52 (2.48, 2.57) & 0.309 (0.298, 0.319) & 49.20 \\
    \addlinespace[3pt]
    0.4 & $-1$ & 1 (1, 1) & 1 (1, 1) & --- & 0.0946 (0.0938, 0.0953) & 0.00 \\
    0.4 & $-0.8$ & 1 (1, 1) & 1 (1, 1) & --- & 0.154 (0.154, 0.155) & 0.00 \\
    0.4 & $-0.6$ & 1 (1, 1) & 1 (1, 1) & --- & 0.245 (0.245, 0.246) & 0.00 \\
    0.4 & $-0.4$ & 1.01 (0.99, 1.03) & 1 (1, 1) & 1.53 (---) & 0.202 (0.175, 0.229) & 0.05 \\
    0.4 & $-0.2$ & 1.5 (1.33, 1.67) & 0.991 (0.988, 0.995) & 1.64 (1.58, 1.71) & 0.375 (0.32, 0.43) & 1.14 \\
    0.4 & $0$ & 3.48 (3.08, 3.88) & 0.952 (0.941, 0.964) & 1.89 (1.85, 1.93) & 0.458 (0.408, 0.507) & 8.38 \\
    0.4 & $+0.2$ & 5.6 (5.11, 6.09) & 0.896 (0.879, 0.912) & 2.09 (2.05, 2.14) & 0.344 (0.313, 0.374) & 17.70 \\
    0.4 & $+0.4$ & 6.15 (5.73, 6.57) & 0.867 (0.849, 0.884) & 2.23 (2.19, 2.28) & 0.316 (0.299, 0.333) & 28.10 \\
    0.4 & $+0.6$ & 6.9 (6.53, 7.27) & 0.798 (0.781, 0.816) & 2.42 (2.37, 2.47) & 0.303 (0.296, 0.31) & 41.10 \\
    0.4 & $+0.8$ & 6.6 (6.2, 7) & 0.776 (0.755, 0.797) & 2.52 (2.46, 2.57) & 0.309 (0.301, 0.317) & 49.50 \\
    0.4 & $+1$ & 6.19 (5.8, 6.58) & 0.732 (0.711, 0.753) & 2.65 (2.59, 2.72) & 0.302 (0.296, 0.309) & 59.50 \\
    \addlinespace[3pt]
    0.5 & $-1$ & 1 (1, 1) & 1 (1, 1) & --- & 0.0901 (0.0894, 0.0909) & 0.00 \\
    0.5 & $-0.8$ & 1 (1, 1) & 1 (1, 1) & --- & 0.158 (0.158, 0.159) & 0.00 \\
    0.5 & $-0.6$ & 1 (1, 1) & 1 (1, 1) & --- & 0.262 (0.261, 0.263) & 0.00 \\
    0.5 & $-0.4$ & 1 (1, 1) & 1 (1, 1) & --- & 0.39 (0.383, 0.397) & 0.31 \\
    0.5 & $-0.2$ & 1.34 (1.23, 1.45) & 0.994 (0.991, 0.996) & 1.49 (1.46, 1.53) & 0.288 (0.248, 0.328) & 0.61 \\
    0.5 & $0$ & 3.77 (3.35, 4.19) & 0.95 (0.94, 0.96) & 1.85 (1.81, 1.88) & 0.434 (0.39, 0.479) & 8.16 \\
    0.5 & $+0.2$ & 6.32 (5.83, 6.81) & 0.87 (0.851, 0.888) & 2.11 (2.07, 2.15) & 0.353 (0.324, 0.382) & 21.50 \\
    0.5 & $+0.4$ & 6.68 (6.32, 7.04) & 0.848 (0.831, 0.865) & 2.28 (2.24, 2.32) & 0.32 (0.304, 0.336) & 34.90 \\
    0.5 & $+0.6$ & 6.89 (6.54, 7.24) & 0.801 (0.782, 0.82) & 2.38 (2.34, 2.42) & 0.31 (0.302, 0.319) & 43.60 \\
    0.5 & $+0.8$ & 6.44 (6.02, 6.86) & 0.742 (0.718, 0.767) & 2.52 (2.47, 2.57) & 0.3 (0.292, 0.308) & 56.10 \\
    0.5 & $+1$ & 6.1 (5.7, 6.5) & 0.718 (0.695, 0.74) & 2.62 (2.55, 2.68) & 0.308 (0.302, 0.313) & 60.70 \\
    \bottomrule
  \end{tabular}
\end{table}

\clearpage

\begin{table}[htbp]
  \centering
  \caption{Cluster summary for $m = 16$}
  \label{tab:m16}
  \footnotesize
  \setlength{\tabcolsep}{4pt}
  \begin{tabular}{cc ccccr}
    \toprule
    $\varepsilon$ & $\gamma$ & $N_{\text{cluster}}$ & $P_{\text{connected}}$ & Inter-cluster distance & Intra-cluster distance & Noise (\%) \\
    \midrule
    0.1 & $-1$ & 1.42 (1.24, 1.6) & 0.997 (0.995, 0.999) & 1.89 (1.81, 1.97) & 0.463 (0.387, 0.538) & 0.43 \\
    0.1 & $-0.8$ & 2.89 (2.4, 3.38) & 0.979 (0.972, 0.986) & 2.04 (1.97, 2.12) & 0.653 (0.58, 0.727) & 1.80 \\
    0.1 & $-0.6$ & 5.3 (4.49, 6.11) & 0.955 (0.944, 0.966) & 2.21 (2.15, 2.28) & 0.735 (0.668, 0.802) & 5.27 \\
    0.1 & $-0.4$ & 7.11 (6.24, 7.98) & 0.929 (0.916, 0.943) & 2.31 (2.26, 2.37) & 0.663 (0.605, 0.721) & 7.75 \\
    0.1 & $-0.2$ & 9.76 (8.65, 10.9) & 0.893 (0.876, 0.91) & 2.47 (2.41, 2.53) & 0.609 (0.557, 0.661) & 14.60 \\
    0.1 & $0$ & 11.8 (10.7, 12.9) & 0.868 (0.849, 0.887) & 2.59 (2.53, 2.65) & 0.543 (0.495, 0.592) & 19.20 \\
    0.1 & $+0.2$ & 13.8 (12.8, 14.7) & 0.842 (0.825, 0.859) & 2.7 (2.66, 2.75) & 0.47 (0.429, 0.512) & 24.60 \\
    0.1 & $+0.4$ & 14.7 (13.8, 15.7) & 0.816 (0.799, 0.833) & 2.81 (2.75, 2.87) & 0.441 (0.401, 0.482) & 32.10 \\
    0.1 & $+0.6$ & 15.1 (14.2, 16.1) & 0.811 (0.794, 0.828) & 2.89 (2.83, 2.94) & 0.431 (0.395, 0.467) & 36.30 \\
    0.1 & $+0.8$ & 15.3 (14.6, 16) & 0.79 (0.773, 0.807) & 2.96 (2.9, 3.01) & 0.362 (0.346, 0.379) & 42.30 \\
    0.1 & $+1$ & 16 (15.3, 16.7) & 0.792 (0.776, 0.808) & 3.11 (3.05, 3.17) & 0.347 (0.331, 0.364) & 49.90 \\
    \addlinespace[3pt]
    0.2 & $-1$ & 1 (1, 1) & 1 (1, 1) & --- & 0.14 (0.139, 0.141) & 0.00 \\
    0.2 & $-0.8$ & 1 (1, 1) & 1 (1, 1) & --- & 0.184 (0.18, 0.188) & 0.00 \\
    0.2 & $-0.6$ & 1.13 (1.05, 1.21) & 0.999 (0.999, 1) & 1.67 (1.62, 1.72) & 0.262 (0.202, 0.322) & 0.12 \\
    0.2 & $-0.4$ & 2.58 (2.13, 3.03) & 0.986 (0.981, 0.991) & 1.95 (1.88, 2.02) & 0.677 (0.606, 0.748) & 1.64 \\
    0.2 & $-0.2$ & 6.15 (5.3, 7) & 0.943 (0.932, 0.955) & 2.19 (2.13, 2.24) & 0.73 (0.672, 0.788) & 6.24 \\
    0.2 & $0$ & 11.6 (10.6, 12.6) & 0.879 (0.864, 0.894) & 2.5 (2.45, 2.55) & 0.51 (0.468, 0.552) & 17.20 \\
    0.2 & $+0.2$ & 14.4 (13.3, 15.5) & 0.832 (0.813, 0.85) & 2.72 (2.66, 2.78) & 0.432 (0.395, 0.469) & 28.00 \\
    0.2 & $+0.4$ & 15.6 (14.8, 16.5) & 0.811 (0.795, 0.827) & 2.93 (2.86, 2.99) & 0.392 (0.363, 0.42) & 40.20 \\
    0.2 & $+0.6$ & 16 (15.2, 16.7) & 0.796 (0.781, 0.811) & 3.07 (3.01, 3.13) & 0.335 (0.324, 0.347) & 50.20 \\
    0.2 & $+0.8$ & 15.4 (14.6, 16.2) & 0.778 (0.762, 0.794) & 3.18 (3.11, 3.24) & 0.336 (0.326, 0.347) & 58.60 \\
    0.2 & $+1$ & 14.1 (13.4, 14.8) & 0.806 (0.79, 0.822) & 3.37 (3.29, 3.44) & 0.332 (0.315, 0.348) & 65.40 \\
    \addlinespace[3pt]
    0.3 & $-1$ & 1 (1, 1) & 1 (1, 1) & --- & 0.135 (0.134, 0.135) & 0.00 \\
    0.3 & $-0.8$ & 1 (1, 1) & 1 (1, 1) & --- & 0.201 (0.2, 0.201) & 0.00 \\
    0.3 & $-0.6$ & 1 (1, 1) & 1 (1, 1) & --- & 0.302 (0.299, 0.306) & 0.00 \\
    0.3 & $-0.4$ & 1.18 (1.07, 1.29) & 0.999 (0.998, 1) & 1.68 (1.6, 1.75) & 0.305 (0.251, 0.359) & 0.14 \\
    0.3 & $-0.2$ & 3.59 (3.06, 4.12) & 0.971 (0.963, 0.979) & 1.91 (1.85, 1.97) & 0.656 (0.591, 0.722) & 2.83 \\
    0.3 & $0$ & 9.56 (8.58, 10.5) & 0.901 (0.885, 0.917) & 2.35 (2.3, 2.39) & 0.615 (0.561, 0.67) & 12.40 \\
    0.3 & $+0.2$ & 15.3 (14.4, 16.3) & 0.829 (0.812, 0.845) & 2.71 (2.65, 2.76) & 0.43 (0.396, 0.463) & 30.80 \\
    0.3 & $+0.4$ & 16.6 (15.7, 17.4) & 0.811 (0.796, 0.826) & 3.02 (2.96, 3.09) & 0.356 (0.336, 0.376) & 50.90 \\
    0.3 & $+0.6$ & 16 (15.3, 16.7) & 0.803 (0.789, 0.817) & 3.21 (3.14, 3.28) & 0.325 (0.316, 0.333) & 60.60 \\
    0.3 & $+0.8$ & 13.7 (12.9, 14.5) & 0.811 (0.794, 0.828) & 3.46 (3.38, 3.54) & 0.308 (0.301, 0.315) & 71.70 \\
    0.3 & $+1$ & 11.2 (10.2, 12.2) & 0.847 (0.83, 0.864) & 3.71 (3.6, 3.81) & 0.299 (0.293, 0.305) & 79.60 \\
    \addlinespace[3pt]
    0.4 & $-1$ & 1 (1, 1) & 1 (1, 1) & --- & 0.13 (0.129, 0.13) & 0.00 \\
    0.4 & $-0.8$ & 1 (1, 1) & 1 (1, 1) & --- & 0.21 (0.209, 0.211) & 0.00 \\
    0.4 & $-0.6$ & 1 (1, 1) & 1 (1, 1) & --- & 0.335 (0.334, 0.335) & 0.01 \\
    0.4 & $-0.4$ & 1 (1, 1) & 1 (1, 1) & --- & 0.284 (0.258, 0.309) & 0.84 \\
    0.4 & $-0.2$ & 2.71 (2.23, 3.19) & 0.985 (0.98, 0.99) & 1.83 (1.77, 1.89) & 0.604 (0.532, 0.677) & 1.78 \\
    0.4 & $0$ & 9.93 (8.86, 11) & 0.897 (0.883, 0.912) & 2.27 (2.21, 2.32) & 0.577 (0.532, 0.622) & 13.50 \\
    0.4 & $+0.2$ & 16.4 (15.6, 17.3) & 0.812 (0.796, 0.828) & 2.7 (2.65, 2.74) & 0.405 (0.379, 0.431) & 34.90 \\
    0.4 & $+0.4$ & 16.8 (16.1, 17.4) & 0.793 (0.778, 0.808) & 3.05 (3, 3.11) & 0.338 (0.329, 0.347) & 57.30 \\
    0.4 & $+0.6$ & 14.5 (13.7, 15.4) & 0.823 (0.805, 0.84) & 3.36 (3.28, 3.43) & 0.314 (0.306, 0.322) & 70.20 \\
    0.4 & $+0.8$ & 12.1 (11.3, 13) & 0.835 (0.816, 0.854) & 3.63 (3.53, 3.72) & 0.303 (0.297, 0.31) & 77.90 \\
    0.4 & $+1$ & 8.3 (7.41, 9.19) & 0.874 (0.855, 0.892) & 3.92 (3.83, 4.02) & 0.29 (0.283, 0.297) & 87.40 \\
    \addlinespace[3pt]
    0.5 & $-1$ & 1 (1, 1) & 1 (1, 1) & --- & 0.125 (0.124, 0.125) & 0.00 \\
    0.5 & $-0.8$ & 1 (1, 1) & 1 (1, 1) & --- & 0.216 (0.215, 0.216) & 0.00 \\
    0.5 & $-0.6$ & 1 (1, 1) & 1 (1, 1) & --- & 0.359 (0.358, 0.359) & 0.04 \\
    0.5 & $-0.4$ & 1.01 (0.99, 1.03) & 1 (1, 1) & 0.411 (---) & 0.51 (0.509, 0.51) & 14.40 \\
    0.5 & $-0.2$ & 1.88 (1.63, 2.13) & 0.993 (0.99, 0.995) & 1.66 (1.6, 1.72) & 0.481 (0.42, 0.542) & 0.75 \\
    0.5 & $0$ & 11.2 (10.3, 12.2) & 0.887 (0.872, 0.902) & 2.28 (2.24, 2.32) & 0.559 (0.516, 0.602) & 15.90 \\
    0.5 & $+0.2$ & 16.4 (15.6, 17.3) & 0.821 (0.806, 0.836) & 2.68 (2.63, 2.72) & 0.391 (0.37, 0.412) & 38.40 \\
    0.5 & $+0.4$ & 15.8 (14.9, 16.6) & 0.817 (0.801, 0.834) & 3.11 (3.03, 3.18) & 0.331 (0.323, 0.34) & 63.50 \\
    0.5 & $+0.6$ & 12.8 (11.9, 13.6) & 0.849 (0.834, 0.865) & 3.56 (3.48, 3.63) & 0.305 (0.3, 0.31) & 78.60 \\
    0.5 & $+0.8$ & 8.85 (8.13, 9.57) & 0.888 (0.871, 0.906) & 4.01 (3.9, 4.11) & 0.289 (0.283, 0.295) & 87.30 \\
    0.5 & $+1$ & 6.59 (5.91, 7.27) & 0.897 (0.87, 0.925) & 4.21 (4.09, 4.33) & 0.279 (0.27, 0.287) & 90.70 \\
    \bottomrule
  \end{tabular}
\end{table}

\end{document}